\documentclass[conference]{IEEEtran}
\IEEEoverridecommandlockouts
\usepackage{amsmath,amssymb,amsfonts}
\usepackage{algorithmic}
\usepackage{graphicx}
\usepackage{textcomp}
\usepackage{xcolor}
\usepackage[ruled,linesnumbered]{algorithm2e}
\usepackage{cite}
\usepackage{multicol}
\usepackage{subfig}

\def\BibTeX{{\rm B\kern-.05em{\sc i\kern-.025em b}\kern-.08em
    T\kern-.1667em\lower.7ex\hbox{E}\kern-.125emX}}
\begin{document}

\title{Adversarial Embedding: A robust and elusive Steganography and Watermarking technique\\
}

\author{\IEEEauthorblockN{Salah Ghamizi\IEEEauthorrefmark{1},
Maxime Cordy\IEEEauthorrefmark{1}, Mike Papadakis\IEEEauthorrefmark{1} and
Yves Le Traon\IEEEauthorrefmark{1}}
\IEEEauthorblockA{SnT, University of Luxembourg}\\
Email: \IEEEauthorrefmark{1}{firstname.lastname}@uni.lu}
\maketitle


\begin{abstract}
We propose \emph{adversarial embedding}, a new steganography and watermarking technique that embeds secret information within images. 
The key idea of our method is to use deep neural networks for image classification and adversarial attacks to embed secret information within images. Thus, we use the attacks to embed an encoding of the message within images and the related deep neural network outputs to extract it. The key properties of adversarial attacks (invisible perturbations, non-transferability, resilience to tampering) offer guarantees regarding the confidentiality and the integrity of the hidden messages. We empirically evaluate \emph{adversarial embedding} using more than 100 models and 1,000 messages. Our results confirm that our embedding passes unnoticed by both humans and steganalysis methods, while at the same time impedes illicit retrieval of the message (less than 13\% recovery rate when the interceptor has some knowledge about our model), and is resilient to soft and (to some extent) aggressive image tampering (up to 100\% recovery rate under jpeg compression). We further develop our method by proposing a new type of adversarial attack which improves the embedding density (amount of hidden information) of our method to up to 10 bits per pixel.



\end{abstract}

\begin{IEEEkeywords}
deep neural network, adversarial attack, image embedding, steganography, watermarking, privacy
\end{IEEEkeywords}

\setcounter{section}{0}
\maketitle
\section{Introduction}
\label{sec:introduction}

Machine Learning (ML) is used to tackle various types of complex problems. For instance, ML can predict events based on the patterns and features they learnt from previous examples (the training set) and to classify given inputs in specific categories (classes) learnt during training. 


Although ML and, in particular, Deep Neural Networks (DNNs) have provided impressive results, they are vulnerable to the so-called adversarial examples. Adversarial examples are inputs that fool ML, i.e., cause misclassifications. The interesting thing with adversarial examples is that the misclassifications are triggered by a systematic procedure. This procedure is tailored to alter the input data (images in particular) in such a way that is not noticeable by human eyes. 


The elusive property of adversarial data has always been perceived by the research community as a major weakness that should be avoided or mitigated.   
While this is important for specific application domains, we argue that such a property can be useful, e.g., to support watermaking and steganography applications. Therefore, our paper takes a different perspective on adversarial attacks by showing that the main properties of targeted attacks (invisibility, non-transferability, resilience and adaptation against input tampering) can be used to form strong watermaking and steganography techniques, which we call Adversarial Embedding. 

Digital watermarking and steganography are methods aiming at hiding secret information in images (or other digital media), by slightly altering original images (named the cover images) to embed the information within them. Watermarking techniques require the embedding to be robust (i.e. resilient to malicious tampering) while steganography focuses on maximizing the amount of embedded data (that is, achieve high-density embedding). 
Of course, undetectability and un-recoverability by a non-authorized third party are of foremost importance in both cases. Thus, the image embedding the secret information (named the stego image) should not be detected by steganalysis techniques.



Our objective is to pave the way for a new generation of watermarking and steganography techniques relying on adversarial ML. We believe that 

Existing steganography and watermarking techniques either are easily detected, manage to embed limited amount of information, or are easily recoverable \cite{stegano_survey}. This means that there are multiple dimensions on which the techniques need to succeed. Notably, previous research achieves relatively good results in a single dimension but not all. In contrast, our technique dominates on multiple dimensions together. For instance, we can embed much more information without risking them to be recovered by a third party. We can also outperform the performance of existing techniques by considering single dimensions alone. All in all, our paper offers a novel and effective watermarking and steganography technique.

Applying our technique requires a DNN classification model with multiple output categories and a targeted adversarial attack. In the case of steganography, we assume that the model has been 'safely' shared between the people that exchange messages. In the case of watermarking, the person who embeds and extracts the messages is the same and thus there is no need for such an assumption. 

The classification model is used to extract the hidden messages by mapping the output classes to bits. The adversarial attack (non-transferable targeted attack on the shared model) is used to embed information by creating adversarial images (the stego images) from original images (the cover images) in a way that the model is forced to classify the adversarial images in the desired classes (corresponding to the message to encode). Since the attack is non-transferable, only the shared model can identify the embedding and give the sought outputs. 

The contributions made by this paper can be summarised in the following points:

We propose a pipeline that uses adversarial attacks to embed information in images. It can be used both for image watermarking and steganography and is founded on the fast-pacing research on adversarial attacks. We call this approach Adversarial Embedding. 
Our pipeline relies on a targeted adversarial attack called \textbf{S}orted \textbf{T}argeted \textbf{A}dversarial \textbf{A}ttack (SATA) that we specifically develop for this work. Our adversarial attack increases the amount of possible data that can be hidden by using multi-class embedding. SATA is capable of embedding seven times more data than existing adversarial attacks with small models (with 10 output classes like the Cifar-10 dataset), and orders of magnitude more with bigger models (with 100 output classes for example).

The steganography literature contains few approaches that use Deep Learning to embed data. Baluja proposed a system with 2 deep neural networks, an encoder and a decoder\cite{Baluja2017HidingII}. Zhu et al. expanded that idea to a system with 3 convolutional neural network to play the Encoder/Decoder/Adversary \cite{zhu2018hidden}. Volkhonskiy et al. on the other hand proposed to use a custom Generative Adversarial Network \cite{gan_stegano} to embed information.  

The main advantage of our pipeline over the previous techniques relying on deep learning is that it does not require a specific model to be designed and trained for this task. It can indeed be used with any image classification model and adversarial attack to enforce higher security.

We demonstrate that our pipeline has competitive steganography properties. The fact that SATA (the new attack we propose) builds upon a state-of-the-art adversarial attack algorithm allows us to generate minimal perturbation on the cover images. This places our approach among the steganography techniques with the least footprint.

We also show that Adversarial Embedding with SATA can achieve almost 2 times the density of the densest steganography technique \cite{stegano_survey}, with an embedding density up to 10 bits per pixel.

We analyze the resilience of our system to tampering and show that, because our system allows the freedom to choose any image classification model in the pipeline, we can find a combination of classification models and adversarial attacks resilient to image tampering with more than 90\% recovery rate under perturbation.

Finally, we assess the secrecy of our system and demonstrate that available steganalysis tools poorly perform in detecting or recovering data hidden with adversarial embedding.

\section{Background and Related Work}
\label{sec:rqs}

\subsection{Steganography}

\begin{figure}
\centering
\includegraphics[width=8cm]{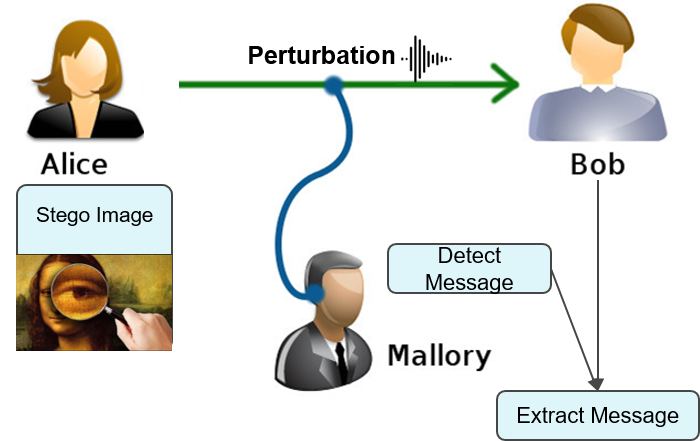}
\caption{\textbf{Sending and decoding a stego-image}: When Alice sends the crafted images to Bob, they can suffer perturbation while in transit. These perturbations can be random or malicious. Besides, a third party would try to detect if the image contains any hidden message, and extract the message if any. Bob, on the other hand, possesses the right decoder to recover the original message.}
\label{fig:atk-decoding}
\end{figure}

Steganography is the process of hiding important information in a trivial medium. It can be used to transmit a message between a sender A and a receiver B in a way that a malicious third party M cannot detect that the medium contains a hidden message and, in the case M still detects the message, M should not be able to extract it from the medium (Fig \ref{fig:atk-decoding}).

The term Steganographia was introduced at the end of the 15th century by Trithemius, who hid relevant messages in his books. Since then, steganography expanded and has been used in various media, such as images \cite{image_stegano} and audio \cite{audio_stegano}. 

Steganography techniques use either the spatial domain or the frequency domain to hide information. 

When operating on the spatial domain, the steganography algorithm changes adaptively some pixels on the image to embed data. Basic techniques to embed messages in the spatial domain include LSB (Least Significant Bit) \cite{lsb_embedding} and PVD (Pixel-value Differencing) \cite{pvd_embedding}. Among last generation spatial steganography techniques, the most popular are:

\begin{itemize}
  \item \textbf{HUGO\cite{HUGO} - Highly Undetectable steGO:}\\ HUGO was designed to hide 7 times longer message than LSB matching with the same level of detectability.
  \item \textbf{WOW\cite{WOW} - Wavelet Obtained Weights:}\\  WOW uses syndrome-trellis codes to minimize the expected distortion for a given payload. 
  \item \textbf{HILL \cite{HILL} -High-pass, Low-pass, and Low-pass:}\\ HILL proposed a new cost-function that can be used to improve existing steganography. It uses a high-pass filter to identify the area of the image that would be best to target (less predictable parts).
  \item \textbf{S-UNIWARD\cite{UNIWARD} - Spatial UNIversal WAvelet Relative Distortion:}\\ UNIWARD is another cost function, it uses the sum of relative changes between the stego-images and cover images. It has the same detectability properties as WOW but is faster to compute and optimize and can also be used in the frequency domain. 
\end{itemize}

Frequency domain steganography, on the other hand, relies on frequency distortions \cite{survey_steganalysis} to generate the perturbation. For instance during the JPEG conversion of the picture. Such distortions include Discrete Cosine Transform (DCT), Discrete Wavelet Transform (DWT) and Singular Value Decomposition(SVD). In a nutshell, the frequency domain attacks change some coefficients of the distortions in a way that can only be detected and decoded by the recipient.

Many techniques have been built around this approach, the most known are J-UNIWARD\cite{UNIWARD} and F5\cite{f5_embedding}.

The technique we propose, \emph{adversarial embedding} uses images as media. Its novelty lies in the use of adversarial attack algorithms that can embed the sought messages in the form of classification results (of adversarial examples) of a given ML model. The information is not embedded in the pixels themselves but in the latent representation of the image formed by the classification (ML) model that processes the image. Some of these latent representations can be tailored using adversarial attacks and be extracted in the shape of classification classes. 

\subsection{Watermarking}

Digital Watermarking aims at hiding messages inside a medium. Unlike Steganography, however, the recipient is not supposed to extract the message. That is, when someone, say A, watermarks a medium and shares it with a receiver, say B, it is expected that neither B nor any third party M can detect the watermark and decode it (only A should be able to do that). Watermarking has multiple applications like copyright protection or tracking and authenticating the sources of the mediums.

Thus, while digital watermarking and steganography use the same embedding techniques, they favour different qualities of the embedding. For instance, steganography aims to maximize the quantity of data that can be embedded within the medium. Watermarking focuses more on the integrity of the embedded message, in particular when the medium is subject to perturbations during its life cycle.

\subsection{Steganalysis}

Steganalysis is the field opposite to steganography. It is dedicated to the detection of messages hidden using steganography techniques.
Traditionally, steganalysis had to manual extract the features relevant to the detection of every Steganography technique.

Steganalysis techniques use various approaches to detect hidden messages\cite{Karampidis2018}. Hereafter some of the most notable approaches when facing spatial-domain steganography:

\begin{itemize}
  \item \textbf{Visual Steganalysis}: It analyzes the pixel values and their distribution. It is sufficient to detect basic inconsistencies, for instance, unbalanced distribution of zeroes and ones that indicates LSB steganography. Given the original cover images, visual steganalysis techniques can identify the difference between the two and assess whether this noise is an artefact or if it is meaningful.
  \item \textbf{Signature Steganalysis:} Many steganography solutions append remarkable patterns at the end of the embedded message. For instance, Hiderman steganography software adds \emph{CDN} at the end while Masker, another tool dedicates, the last 77 bytes of the stego-image for its signature. A steganalysis tool would scan the files for such signatures.  
  \item \textbf{Statistical steganalysis:} Detectors of this family focus on some statistics that are commonly modified as a result of the embedding process. SPA (Sample Pair Analysis) method\cite{spa_detector}, RS (Regular and Singular groups) method\cite{rs_detector} and DIH (Difference Image Histogram) method\cite{dih_detector} are among the most popular statistical steganalysis techniques.
  \item \textbf{Deep Learning steganalysis:} Deep learning models are becoming more popular as steganalysis approach. They can be trained to learn the features of existing steganography approaches and detecting them with high rates\cite{survey_steganalysis}. They require however the interception of a large amount of cover and stego-images and the ability to label these images \cite{deeplearning_steganalysis}.
\end{itemize}

\subsection{Adversarial examples}

Adversarial examples result from applying intentional small perturbation to original inputs to alter the prediction of an ML model. In classification tasks, the effect of the perturbation goes from reducing the confidence of the model to making it misclassify the adversarial examples. Seminal papers on adversarial examples \cite{biggio_atk, Szegedy2013IntriguingPO, Goodfellow2014} consider adversarial examples as a security threat for ML models and provide algorithms (commonly named ``adversarial attacks'') to produce such examples.

Since these findings, researchers have played a cat-and-mouse game. On the one hand, they design defence mechanisms to make the ML model robust against adversarial attacks (avoiding misclassifications), such as distillation \cite{Papernot2016}, adversarial training \cite{Kurakin2016AdversarialML}, generative adversarial networks \cite{Samangouei2018DefenseGANPC} etc. On the other hand, they elaborate stronger attack algorithms to circumvent these defences (e.g., PGD \cite{Madry2017}, CW \cite{Carlini2017}).

One can categorise the adversarial attack algorithms in three general categories, black-box, grey-box and white-box. Black-box algorithms assume no prior knowledge about the model, its training set and defence mechanisms. Grey-box ones hold partial knowledge about the model and the training set but have no information regarding the defence mechanisms. Finally, white-box algorithms have full knowledge about the model, its training set and defence mechanisms.

The literature related to applications of adversarial examples is scarcer and mainly rely on the ability of those examples to fool ML-based systems.focuses on their ability to fool ML-based systems used for, e.g., image recognition \cite{face_reco_2016}, malware detection \cite{Anderson2017EvadingML} or porn filtering \cite{Yuan2019}. In this work, we rather consider adversarial examples as a mean of embedding and hiding secret messages.

Our embedding approach relies on 3 attributes of adversarial examples:

\begin{itemize}
  
  \item \textbf{Universal existence of adversarial examples:} Given sufficient amount of noise $\epsilon$, we can always craft an adversarial image for a model $M_1$, starting from any cover image $I$ to be classified into target class $v1$. 
  \item \textbf{Non-transferability of targeted adversarial examples:} A targeted adversarial image crafted for a model $M_1$ to be classified as $v1$ will have a low probability to be classified as $v1$ if we use any model $M_2, \: M_2 != M_1$ to decode it.
  \item \textbf{Resilience to tampering of adversarial examples:} A targeted adversarial image crafted for a model $M_1$ to be classified as $v1$ will still be classified as $v1$ by $M_1$ if the image has suffered low to average tampering and perturbation.
  
\end{itemize}

These attributes have been studied by previous research on adversarial examples \cite{Papernot2016,liu2016delving,Carlini2017,Athalye2018} and we will show in the following sections how they make our Adversarial Embedding approach among the best watermarking and steganography techniques. \\


Another active field of research about adversarial examples is the detection of these perturbations. It can be used as a preemptive defence by flagging suspicious inputs to be handled by a specific pipeline (human intervention, complex analysis....).
In 2017, Carlini et al. surveyed 10 popular detection techniques \cite{CarliniDetection2017} and demonstrated that they can all be defeated by adjusting the attack algorithm with a custom loss function. Their work showed that the properties that researchers believed were inherent to the adversarial attack process, and were used to detect them are not a fatality and can be bypassed by simple optimizations in the attack algorithm.

Since his work, new detection mechanisms have been proposed, borrowing techniques from statistical testing \cite{Grosse2017}, information theory \cite{Zhao2018} and even steganography detection \cite{Liu2018}. However, no technique has demonstrated its ability to apply to every model, dataset and attack available in the literature without being easy to bypass by the optimization of existing attack algorithms.


\section{Adversarial Embedding}
\label{sec:adv_embed}

Our objective is to provide an integrated pipeline for image watermarking and steganography that relies on the known properties of adversarial examples. We name our approach \emph{adversarial embedding}. Its principle is to encode a message as a sequence of adversarial images. Each of those images results from altering given original images (cover images) to force a given $N$-class ML model to classify them into a targeted class. 
To decode the message, the recipient has to use the same model to classify the adversarial images to retrieve their class numbers, which together form the decoded message.

\subsection{Inputs and Parameters}

More precisely, our pipeline comprises the following components: 

\begin{itemize}
  
    \item \textbf{An image classification dataset $\{I_{i} \}$:} The dataset defines the number $N$ of classes (of the related classification problem) which, in turn, determines the size of the alphabet used for the encoding. We can provide our system with any number of images. Still, adversarial attack algorithms may fail to work on some images. The more images in the dataset, the more we increase the size and diversity of the space of the adversarial images that can be used for the encoding.
    
    \item \textbf{A pair of encoder/decoder $E_\lambda$/$E'_\lambda$:} Having defined $N$ through the choice of the dataset, we have to transform a (e.g.) binary secret message in Base $N$. Conversely, at the decoding stage, we transform back the retrieved Base-$N$ message into the original binary message. We use such encoder/decoder pair in a black-box way and make no assumption regarding their properties. For instance, the encoding can be improved to support redundancy or sanity check techniques but this is not a stringent requirement for our approach to work. 
    
    \item \textbf{A classification model $M_\theta$:} Although our method is compatible with any $N$-class classification model that can deal with the chosen dataset, we focus more particularly on Deep Neural Networks (DNNs) as those are arguably the established solution for image classification. $\theta$ denotes the set of parameters that define the model. In DNN classifiers, the parameters include the architecture (number, types and inner parameters of the layers), the weights of the layers learned during training, hyperparameters, etc. The model acts as a secret key for both encoding and decoding. Our approach assumes that the model can be transmitted from the sender to the intended recipient of the message without being intercepted or altered (e.g. through a secure physical device). The choice of the model also impacts the success rate of the adversarial attack algorithms and, thus, the images that can be used for the encoding.
    
    \item \textbf{An adversarial attack algorithm $A_\epsilon$:} We can use any targeted adversarial attack algorithm that can force a model to classify a given image in a chosen class. The hyperparameters of the attack, noted $\epsilon$, include the maximum amount of perturbation allowed on images as well as attack-specific parameters. The choice of the attack algorithm and its parameters impacts the success rate of the attack and the detectability of the perturbation. In this work, we focus more particularly on the Projected Gradient Descent (PGD) \cite{Madry2017} attack because we found it provides a good balance between success rate and perturbation minimization. Moreover, PGD randomly selects starting points around the original images, which makes the embedding non-deterministic (i.e. the same message and the same original image can lead to different steganography images) and, thus, harder to detect (see Section~\ref{sec:results-detection}). Finally, PGD is known to have a low transferability from one model to another \cite{liu2016delving}, which increases the resilience of our approach to illegitimate decoding (see Section~\ref{sec:results-decoding-third-party}). 

\end{itemize}

\subsection{Embedding Pipeline}

\begin{figure}
     \centering
     \subfloat[\textbf{Step 1 of the encoding process}: To embed a 52 bits secret message in a 32x32 color picture. We first select a model that supports image classification of 32x32x3 pictures. We can use a model trained to classify cifar-10 dataset into 10 classes for instance. We encode therefore our message into base 10.]{\includegraphics[width=\linewidth]{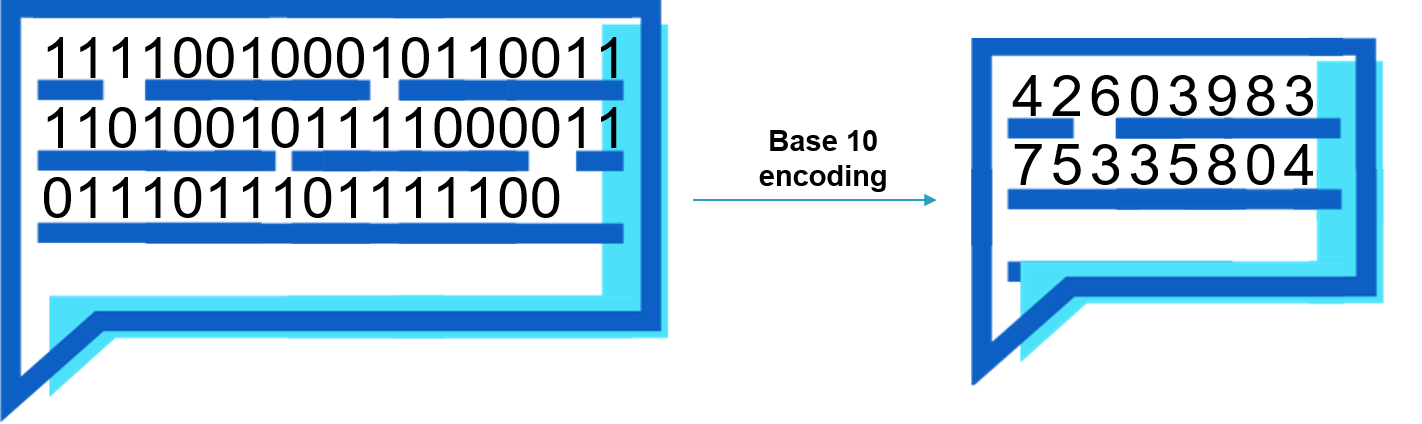} \label{fig:base-encoding}}\\
     
     \subfloat[\textbf{Step 2 of the encoding process}: We convert every element of the encoded message into a logit (a 10 element vector with one 1 and 0 everywhere else). This logit is then used to craft targeted attacks using a cover image (it can be the same or a different cover image per stego-image). Each logit leads to a different stego-image.]{\includegraphics[width=\linewidth]{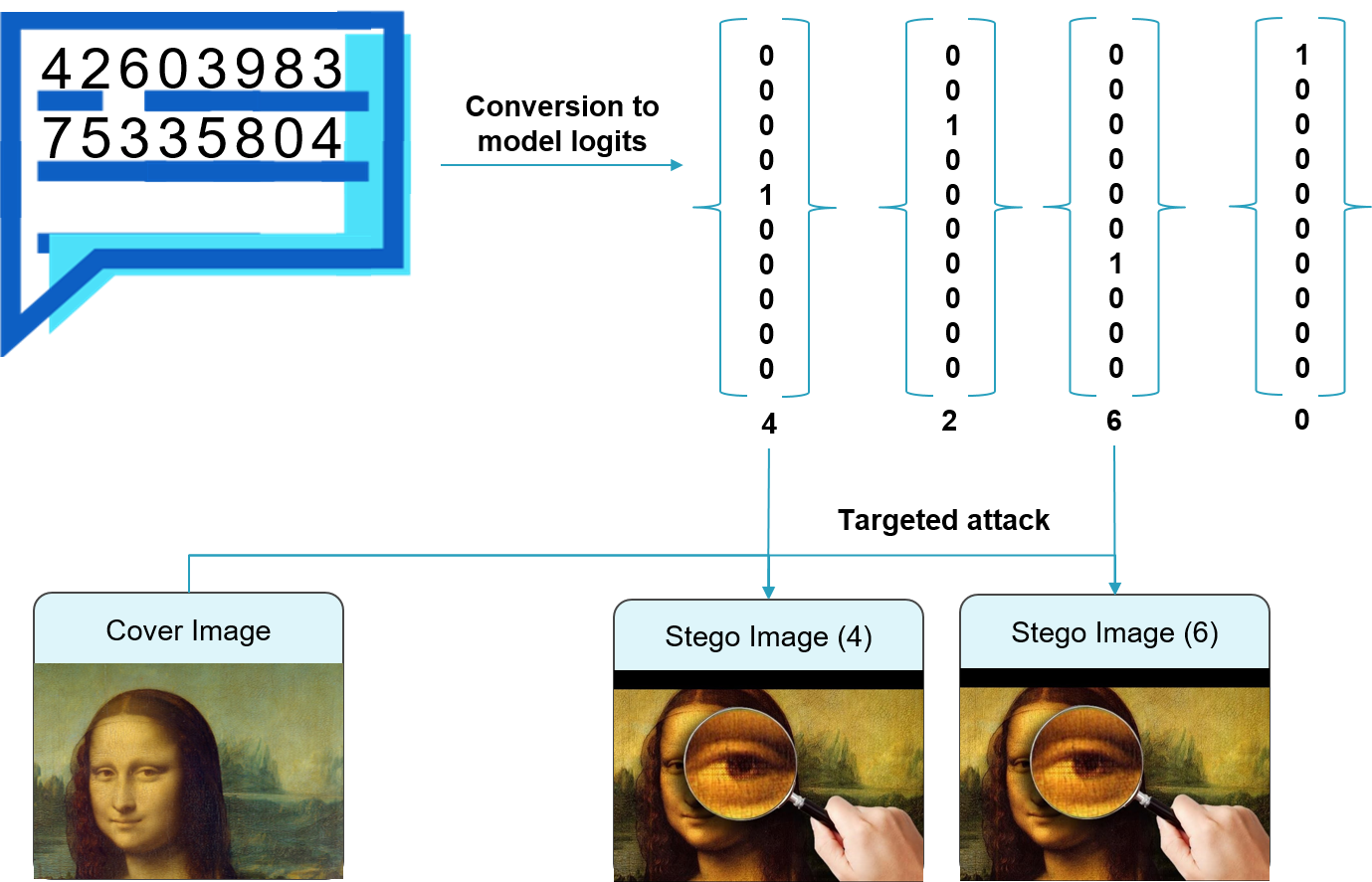} \label{fig:atk-encoding}}
     
    \caption{Embedding a message using targeted adversarial attack.}
     \label{fig:embedding_example}
\end{figure}

Figure \ref{fig:embedding_example} illustrates an instantiation of our approach to embed binary messages $D_{sec} = b_1 \dots b_L \in \{0,1\}^L$ of length $L$ into an image dataset with $N = 10$ classes. First, we use $E_\lambda$ to encode $D_{sec}$ into base 10, resulting in a new message $D_{enc} = n_1 \dots n_{L'} \in \{1..10\}^{L'}$ of length $L' = \lceil L \times \frac{\log 2}{\log N} \rceil$. $L'$ is also the number of adversarial images needed to encode the message. In the second step, we apply the adversarial attack $A_\epsilon$ to insert targeted adversarial perturbations into $L'$ original images $I_1 \dots I_{L'}$ (picked with or without replacement), resulting into $L'$ adversarial images  $A_\epsilon(I_{1}) \dots A_\epsilon(I_{L'})$ such that $M_\theta$ classifies $A_\epsilon(I_{j})$ into class $n_j$. These adversarial images form the sequence of steganography images that are sent to the recipient throughout a (potentially unsecure) channel. While a malicious third party can intercept the sent images, we assume that either the channel is reliable enough to preserve the image ordering or that some consistency mechanism allows to retrieve this ordering reliably.


\subsection{Decoding Pipeline}

Once the recipient receives the adversarial images, he can input them sequentially into the classification $M_\theta$ (which was previously transmitted to her securely) to retrieve their associated class. The resulting class numbers form back the $D_{enc}$ message (in Base 10), which can then go through the decoder $E_{\lambda'}$ to retrieve the original binary message $D_{sec}$.

External disruptions may alter the images, due to the natural process of the carrier (e.g. websites and mobile applications often show only compressed versions of the images) or malicious intent from a third party. For instance, copyright enforcement is a popular application of watermarking. To circumvent this protection while permitting the illegal use of the protected material, malicious people can degrade the mediums with local transformations (e.g. rotation, cropping ...). In such cases, the recipient needs $M_\theta$ to correctly classify altered images resulting from applying the aforementioned transformations to the adversarial images. It is desirable for adversarial embedding to be resilient to such transformation, such that the classification of $M_\theta$ remains preserved in spite of the transformations.

%
%

\section{Sorted Adversarial Targeted Attack (SATA)}


A drawback of adversarial embedding as described previously is that it can only encode $\log_2 N$ bits per adversarial image (where $N$ is the number of classes). In the particular case of Cifar-10 and its 32x32 images and 3 channels, this yields a density of 9.77e-4 bits per pixel (BPP). By contrast, alternative solutions achieve a BPP between 0.1 and 0.2 \cite{zhu2018hidden}. Comparing embedding density is not always fair, though, as the density of adversarial embedding depends on the number of classes and not the number of pixels.\footnote{Established benchmark datasets like MNIST, Cifar-10 and ImageNet exhibit some correlation between their number of classes and image resolution. However, others like NIH Chest X-Ray have large images categorized in few classes.} Thus, its BPP density depends exclusively on the used dataset.

Nevertheless, this limited density originates from the fact that existing adversarial attack algorithms maximize the classification probability of one class, without considering the other classes. This restricts the embedding to use only one value per image, the most probable class value.

To improve density without changing dataset, we propose a new targeted attack that forces the top-$k$ classes suggested by the model for a given image. Thus, the encoding of a message chunk is not only the top class suggested by the model but the top-$k$ ones. We name this type of adversarial attack algorithm \emph{Sorted Adversarial Targeted Attack} (SATA) and we instantiate its principle to extend the PGD algorithm.

Existing targeted adversarial attack algorithms that rely on gradient back-propagation measure the gradient to the target class. This gradient is iteratively used to drive the perturbation added to the input image until the image is misclassified as expected. In SATA, we consider all classes and measure the gradient toward each class. Then, we add a perturbation to the input with a weighted combination of these gradients. The weight is 1 for the top-1 class we want to force, and it decreases gradually for each next class. The more $k$ classes we force, the harder it is to ensure that the classes are sufficiently distinguishable and the harder it is to build an appropriate perturbation.


Apart from that, the remaining parts of our SATA algorithm are similar to PGD: we apply small perturbation steps iteratively until we reach a maximum perturbation value (based on L2 distance) and we repeat the whole process multiple times to randomly explore the space around the original input.

\subsection{Algorithm}

Algorithm \ref{algorithm:encoding_algorithm} describes our SATA algorithm. It calls several internal procedures. \texttt{SplitIntoChunks} (Line 1) takes as input the full message and splits it into chunks, each of which contains up to $k$ different Base-$N$ digits. Each of these numbers represents a class that we want to force into the top $k$ (the $i$-th digit corresponding to the $i$-th top class). The digits must be different because a class cannot occur in two ranks at the same time. This implies that a chunk cannot contain two identical digits. To avoid this, we cut chunks as soon as a digit occurs in it for the second time or when reaching $k$ digits. Then, the number of chunks (into which the full message was decomposed) corresponds to the number of images required to encode the message. For instance, splitting 29234652 into chunks of maximum size $k=3$ using a dataset of $N=100$ classes leads to two chunks $[\{29,23,46\},\{52\}]$. This means that encoding this message requires two adversarial images. The model should classify the first image into 29 as the most probable class, 23 as the second most and 46 as the third most. If we set $k=4$ instead, we can encode the message using one adversarial image, such that the model predicts 52 as the fourth most probable class.


Next, the \texttt{buildWeightedLogits} procedure (Line 2) computes, for each chunk, the weights applied to the gradient of each class when building the perturbation. We start with $w_1 = 1$ and decrease the subsequent weights following an exponential law: $w_i = (1-\frac{(i-1)}{k})^\gamma \:\: \forall{i} \in \{1,...,k\}$.
$\gamma$ is a hyperparameter of SATA that has to be set empirically.

Having our chunks and weights, we enter a loop where we perform $R$ trials to generate a successful perturbation (Lines 5--16). At each trial, \texttt{randomPick} procedure (Line 6) randomly picks, from the whole set of original images, as many images as the number of chunks that must be encoded. 

The procedure \texttt{computeAdv} (Line 9) computes the weighted gradients with all the target classes and use them to drive the adversarial images with a small perturbation step without exceeding the maximum amount of perturbation allowed.

This procedure is elaborated in Algorithm \ref{algorithm:computeAdv_algorithm}. It uses similar sub-procedures as PGD:

\begin{itemize}
  
\item \texttt{RandomSphere(x, epsilon)}: builds a matrix of the same shape as 'x' and a radius epsilon centered around 0. 
\item \texttt{Project(v,eps)}: projects the values in 'v' on the $L_2$ norm ball of size 'eps'.
\item \texttt{LossGradient(x,y)}: measures the loss of the classification model on every input x to its associated classes y.
\end{itemize}

Finally, the \texttt{computeSuccess} procedure (Line 10) checks if the most probable classes are correctly predicted. Any chunk that has one or more classes disordered is considered as a failure (returns 0). If a message requires multiple chunks to be embedded (i.e. multiple cover images), \emph{computeSuccess} returns the average success rate over all chunks.

\IncMargin{1em}
\begin{algorithm}
    \SetKwData{D}{$D_{chunk}$}
    \SetKwData{grpClass}{$weightedLogits$}
    \SetKwData{advX}{$advX$}
    \SetKwData{rate}{$sucessRate$}
    \SetKwData{bestRate}{$bestRate$}
    \SetKwData{bestAdv}{$bestAdv$}
    \SetKwData{bestAdvs}{$bestAdvList$}
    \SetKwData{Istart}{$I_{start}$}
    
    \SetKwFunction{weightedLogits}{buildWeightedLogits}
    \SetKwFunction{SplitMessage}{SplitIntoChunks}
    \SetKwFunction{computeAdv}{computeAdv}
    \SetKwFunction{computeSuccess}{computeSuccess}
    \SetKwFunction{randomPick}{randomPick}
    
    \SetKwInOut{Input}{input}\SetKwInOut{Output}{output}
    \Input{A classifier $M_\theta$; A dataset of cover images $I$ and encoded message $\{D_{enc,i}\}_{i=1}^{L'}$; step size $\epsilon_{step}$; maximum perturbation size $\epsilon$; total iterations $L$; Number of random starts $R$; Number of classes of the model $N$. Number of classes to encode per image $k$ where $k \leq $N }
    \Output{\bestAdv: The stegano-images that encode $D_{enc}$}
    \BlankLine
    
    \D $\leftarrow$ \SplitMessage{$D_{enc}$,$N$,$C$} \\
    \grpClass $\leftarrow$ \weightedLogits{\D}; \\
        \bestRate $\leftarrow 0$ \;
        \bestAdv $\leftarrow Null$ \;
         \For{$j \leftarrow 1$ \KwTo $R$}{
          $I_{start}$ $\leftarrow $\randomPick{$I$,$\|$\D$\|$}\;
          \advX $\leftarrow \Istart$ \;
          \For{$i \leftarrow 1$ \KwTo $L$}{
                \advX $\leftarrow$ \computeAdv{\advX, \grpClass, \Istart, \D, $\epsilon_{step}$, $\epsilon$} ;\\
                \rate $\leftarrow$ \computeSuccess{$M_\theta$, \advX, \grpClass, $m$}; \\         
                  \If{\rate $>$ \bestRate}{
                       \bestRate $\leftarrow$ \rate; \\
                       \bestAdv $\leftarrow$ \advX; \\
                   }
          }
         }
    \caption{SATA algorithm}
    \label{algorithm:encoding_algorithm}
\end{algorithm}
\DecMargin{1em}

\IncMargin{1em}
\begin{algorithm}
    \SetKwData{D}{$D_{chunk}$}
    \SetKwData{grpClass}{$weightedLogits$}
    \SetKwData{advX}{$advX$}
    \SetKwData{grad}{$grad$}
    \SetKwData{perturbation}{$perturbation$}
    \SetKwData{Istart}{$I_{start}$}
    
    \SetKwFunction{RandomSphere}{RandomSphere}
    \SetKwFunction{LossGradient}{LossGradient}
    \SetKwFunction{Project}{Project}
    
    \SetKwInOut{Input}{input}\SetKwInOut{Output}{output}
    \Input{A set of adversarial images \advX; A matrix of weights \grpClass; A set of cover-images \Istart; $y$ A set of targeted classes per image; A maximum perturbation size $\epsilon$; a step perturbation size $\epsilon_{step}$ }
    \Output{\advX: A more perturbed set of adversarial images}
    \BlankLine
    
    \advX $\leftarrow$ $\advX+$ \RandomSphere{\advX,$\epsilon$} \\
    \grad $\leftarrow -1 \times$ \LossGradient{\advX,$y$} \\
    \grad $\leftarrow$ $\frac{\grad}{\|\grad\|}$ \\
    \advX $\leftarrow \advX + \epsilon_{step} \times \grad \cdot \grpClass$ \\
    
    \perturbation $\leftarrow$ \Project{\advX - \Istart,$\epsilon$}\\
    \advX $\leftarrow$ $\Istart + \perturbation$
    \caption{\emph{computeAdv} procedure}
    \label{algorithm:computeAdv_algorithm}
\end{algorithm}
\DecMargin{1em}

\subsection{Embedding Capacity and Density}

For a dataset with $N$ class, the worst-case embedding capacity (in bits) of the standard adversarial embedding (using PGD to encode one digit per image) is $\frac{\log N}{\log 2}$. Thus, with Cifar-10, each image can encode 3 bits. With SATA, each image encodes $k$ classes. 

Let us assume $k = 2$. In this case, each image can encode one of the 90 couples of different 0--9 digits. The couples with the same two digits cannot be encoded since a class cannot be the first and second most probable class at the same time. Thus, in $90\%$ of the cases, the capacity of an image becomes 6 bits. However, when two successive numbers to encode are identical, we have to use two images (each of which has, therefore, a capacity of 3 bits). On average, the embedding capacity of SATA with $k = 2$ and a 10-class dataset is, thus, $6 \times 90\% + 3 \times 10\% = 5.7$ bits. Given that Cifar-10 images contain 32x32 pixels and 3 channels, this yields an embedding density of $1.86e-3$ BPP.

We extend this study empirically to classifiers with 100 classes, 1,000 classses and 10,000 classes in Section \ref{subsection:embedding_capacity_experiments}. 

Further raising $k$ can increase significantly the capacity (although the marginal increase lowers as $k$ goes higher). However, as the adversarial attack has to force the model to rank more classes, its success rate decreases. We investigate this trade-off in our empirical evaluation.

\section{Evaluation Setup}
\label{sec:implementation}

\subsection{Objectives}

Our evaluation aims at determining whether adversarial embedding is a viable technique for steganography and watermarking. For such techniques to work, a first critical requirement is that a third-party should not detect the embedding in the images. Thus, we focus on the ability of adversarial embedding to avoid detection, either by manual or automated inspection. For the first case (manual inspection) we want to ensure that humans cannot identify the adversarial embedding. To validate we check whether the images with the embedding can be distinguished from the original ones. If not, then we can be sure that our embedding can pass unnoticed.  In steganography, this is important for the confidentiality of the message; in watermarking, this ensures that the watermark does not alter the perception of the image by the end-user. Thus, we ask:

\begin{description}
    \item[\textbf{RQ1.}] \hspace{0.05cm} \emph{Does adversarial embedding produce elusive (wrt human perception) images?}
\end{description}

As for automated methods, steganalysis is the field of research that studies the detection of hidden messages and watermarks implemented using steganography techniques. Having intercepted a load of exchanged data, steganalysis methods analyze the data and conclude whether or not they embed hidden messages. We confront adversarial embedding to those methods and ask:

\begin{description}
    \item[\textbf{RQ2}] \hspace{0.05cm} \emph{Can adversarial embedding remain undetected by state-of-the-art steganalysis methods?}
\end{description}

A second security requirement is that, in the case where a third-party detected an embedded message in the cover, it is unable to extract it. In our method, the classification model from which the adversarial embedding is crafted is the key required to decode the message. If we assume that the third-party has no access to this model, the only way to decode is by building a surrogate model and classify the adversarial examples exactly as the key model does. Thus, we ask:

\begin{description}
    \item[\textbf{RQ3.}] \hspace{0.05cm} \emph{Can adversarial embedding be extracted by different models?}
\end{description}

After studying the confidentiality of the embedded information, we turn our attention towards its integrity when reaching the recipient (in steganography) or when checking the authenticity of the source (in watermarking). Integrity is threatened by image tampering. Under the assumption that the model used to craft the adversarial embedding is unaltered, we want to ensure that decoding tampered images still yields the original message. We consider \emph{spatial domain} tampering resulting from basic image transformations (rotation, upscaling and cropping) as well as \emph{frequency domain} tampering like JPEG compression and color depth reduction. We ask:

\begin{description}
    \item[\textbf{RQ4}] \hspace{0.05cm} \emph{Is adversarial embedding resilient to spatial domain and frequency domain tampering?}
\end{description}

The last part of our study focuses on the steganography use case and considers the benefits of our SATA algorithm to increase the density (in bits per pixel) achieved by the adversarial embedding (by targeting $k$ classes). We study this because it is possible to get a smaller success rate by SATA (that considers multiple classes) compared to PGD (which targets a single class). We study this trade-off and ask:

\begin{description}
    \item[\textbf{RQ5}] \hspace{0.05cm} \emph{What are the embedding density and the success rate achieved by of Sorted Adversarial Targeted Attack?}
\end{description}

\noindent Density and success rate are dependent on the number $N$ of classes that are targeted by the attack algorithm. Thus, we study this question for different values of $N$. In particular, we are interested in the maximum value of $N$, which suggest the maximum capacity of SATA to successfully embed information. 

\subsection{Experiment subjects}

\textbf{Messages}. Most of our experiments consider two messages to encode. \texttt{Message1} is a \emph{hello} message, encoded in Base 10 as 29234652. \texttt{Message2} is a randomly-generate message of 100 alpha-numerical characters encoded in Base 10. 

In our RQ5 experiments we asses the embedding density of 3 classifiers with 1000 thousand randomly generated messages each (of length of ~6.64 Kbits for the first 2 classifiers and ~33 Kbits for the third classifier) then use \texttt{Message3}, a randomly-generated message of 6.64 kbits to assess the tradeoff between density and success rate.

\textbf{Image dataset}. We use the Cifar-10 dataset \cite{cifar10} as original images to craft the adversarial embedding. Cifar-10 comprises 60,000 labelled images scattered in 10 classes, with a size of 32x32 pixels and 3 color channels (which make it suitable for watermarking). With 10 classesand one-class embedding (PGD for instance), every image can embed 3 bits ($2^3 \leq 10<2^4$). With the 32x32 pixel size and the 3 color channels, this yields an embedding density of $9.77e^{-4}$ BPP. By comparison, ImageNet usually uses images of 256x256 pixels and supports 21K categories which allow us to embed up to 14bits per image. However, due to the size of images, the image density is only 7.12e-5 BPP. Moreover, classification models for Cifar-10 require reasonable computation resources (compared to ImageNet) to be trained.

\textbf{Classification models}. Our experiments involve two pre-trained models taken from the literature and 100 generated models. The two pre-trained models are (i) the default Keras architecture for Cifar-10\footnote{https://keras.io/examples/cifar10\_cnn/} -- named KerasNet -- and a Resnet 20 (V1) architecture. Both achieve $+80\%$ accuracy with Adam Optimizer, data-augmentation and 50 training epochs. The remaining 100 models were produced by FeatureNet \cite{Ghamizi2019}, a neural architecture search tool that can generate a predefined number of models while maximizing diversity. 



\subsection{Implementation and Infrastructure}

All our experiments run on Python 3.6. We use the popular Keras Framework on top of Tensorflow and various third-party libraries. The Github repository of the project\footnote{https://github.com/yamizi/Adversarial-Embedding} defines the requirements and versions of each library.

To craft the adversarial examples, we use the PGD algorithm with its default perturbation step and maximum perturbation parameters \cite{Madry2017}. To make sure these algorithm and parameters are relevant, we measure the empirical success rate of the algorithm in applying adversarial embedding on Cifar-10 with the Keras model. We encode \texttt{Message 2} into 154 adversarial images produced by applying PGD on 154 original images selected randomly (without replacement). We repeat the process 100 times (resulting in 15,400 runs of PGD) and measure the percentage of times that PGD successfully crafted an adversarial example. We obtain a 99\% success rate, which tends to confirm the relevance of PGD for adversarial embedding.


Model generation and training (using FeatureNet) were performed on a Tesla V100-SXM2-16GB GPU on an HPC node. All remaining experiments were performed on a Quadro P3200-6GB GPU on a I7-8750, 16Gb RAM laptop. 
\section{Evaluation Results}
\label{sec:evaluation}

\subsection{RQ1: Visual Perception of Perturbations}

We rely on two metrics to quantify the perception of the perturbation on adversarial images. The first is Structural Similarlity Index Metric (SSIM) \cite{SSIM2004}, which roughly measures how close two images are. It is known to be a better metric than others like signal-to-noise ratio (PSNR) and mean squared error (MSE). In \cite{SSIM_threshold}, the authors show that humans cannot perceive perturbations with a SSIM loss smaller than 4\%. Depending on the case, humans might start perceiving small perturbations from 4\% to 8\% of SSIM loss.

The second metric is LPIPS \cite{Zhang2018}. It relies on a surrogate model (in our case, we used an AlexNet \cite{alexnet}) to assess perceptual similarity. A preliminary study \cite{Zhang2018} showed that it outperforms SSIM in terms of correlation with human perception and it can compare more finely perturbations with close SSIM values.  

To evaluate whether humans can perceive the perturbations incurred by adversarial embedding, we embed \texttt{Message1} using 8 cover images and \texttt{Message2} using 154 cover images. In both cases, we applied PGD on the KerasNet model to generate the adversarial images. Then, we measure the SSIM and LPIPS losses between the original images and the perturbed images. 
To complement our analysis, we compare the degree of perceived perturbation (as computed by the metrics) with the one resulting from applying a 75\% quality JPEG compression (resulting in a loss of information of 25\%) on the original images.

\begin{table}
\centering
\caption{Perceptual similarity loss between original images and the images produced by adversarial embeddings. The first two lines relate to the 8 original images used to embed \texttt{Message1}. The last two lines relate to the 154 images used to embed \texttt{Message2}.}
\begin{tabular}{r|r|r|r}
Image set & Perturbation  &  SSIM loss (\%) &  LPIPS loss (\%) \\
\hline
\texttt{Message1} & Adv. Emb. & 8.98 +/- 0.75 & 0.29 +/- 3.71e-04 \\
\texttt{Message1} & JPEG (75\%) & 8.17 +/- 0.21 & 1.07 +/-3.44e-03 \\
\texttt{Message2} & Adv. Emb. & 6.02 +/- 0.58 & 0.33 +/- 9.68e-04 \\
\texttt{Message2} & JPEG (75\%) & 6.65 +/- 0.20 & 1.05 +/- 8.44e-03 \\
\end{tabular}

\label{table:comparison_perceptual_similarity}
\end{table}

\begin{figure}
\centering
\subfloat[Mean SSIM loss values between the original cover image and the stego image. The lower the better.]{\includegraphics[width=\linewidth,trim={4cm 1cm 4cm 3cm},clip]{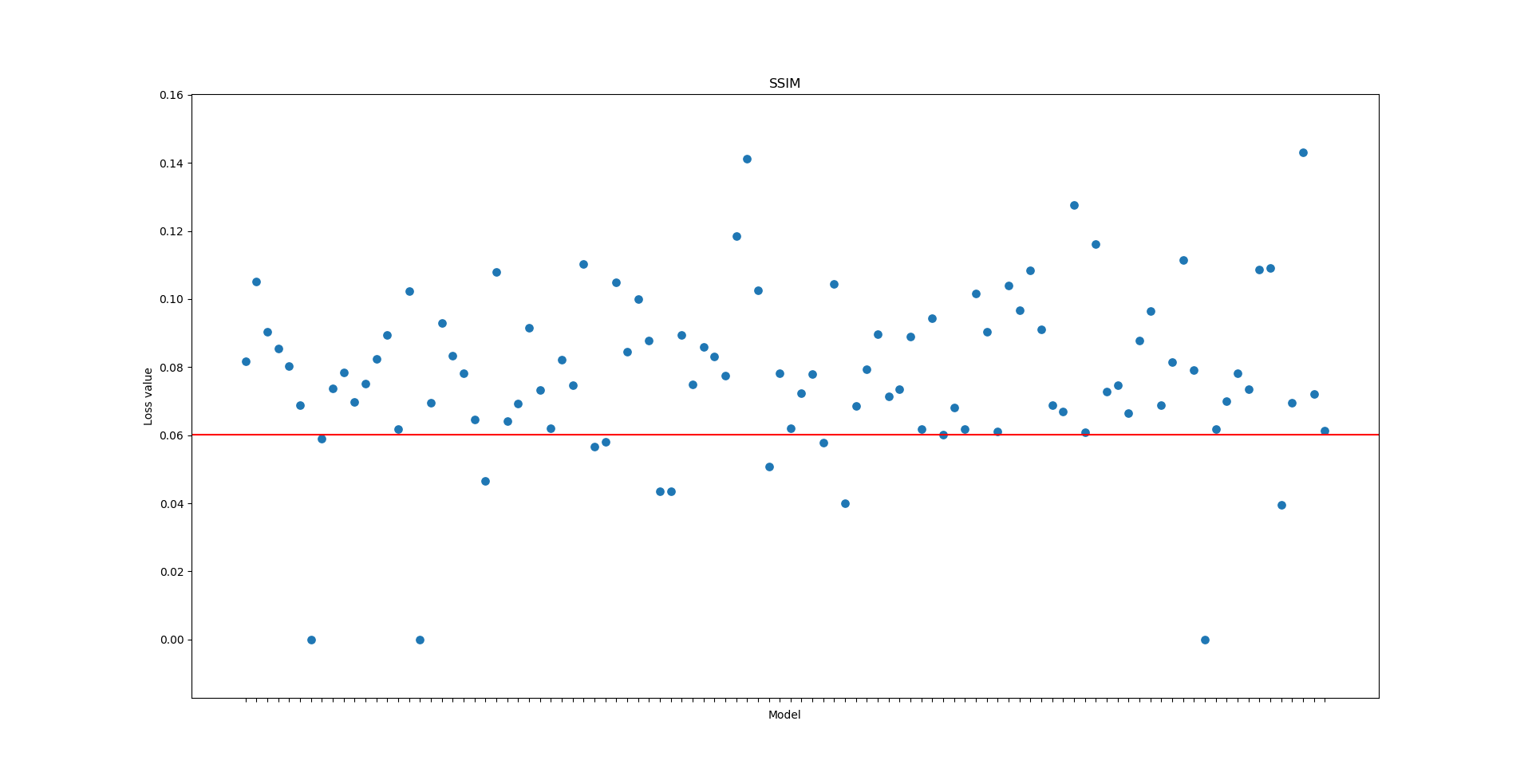}} \hfill
\subfloat[Mean LPIPS loss values between the original cover image and the stego image. The lower the better.]{\includegraphics[width=\linewidth,trim={4cm 1cm 4cm 3cm},clip]{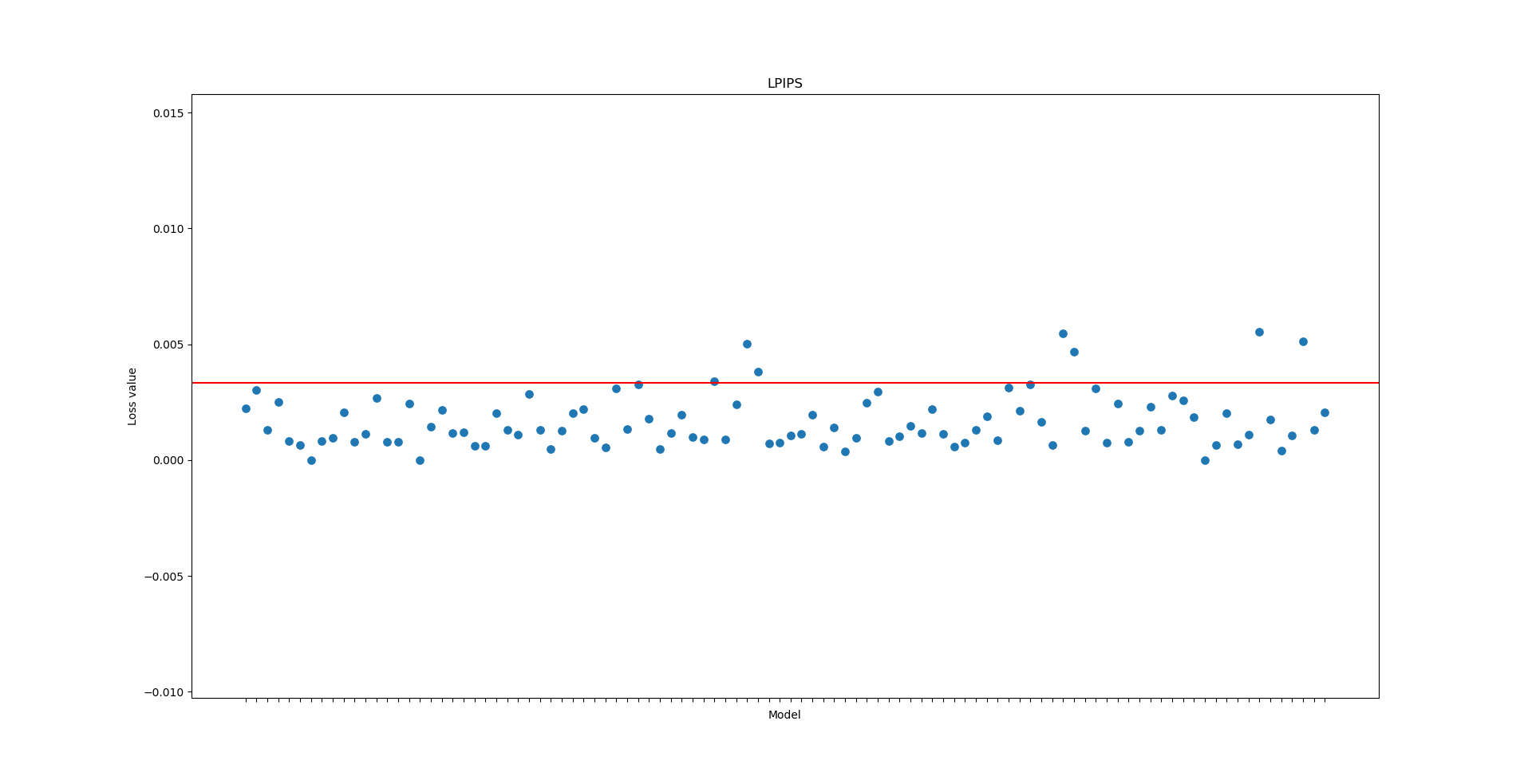}}
\caption{SSIM \& LPIPS loss values of the images encoded using one of the 100 sampled models. The red lines indicate the values we obtained by the KerasNet model we studied in details in Table \ref{table:comparison_perceptual_similarity}.}
\label{fig:sampling_similarity}
\end{figure}

Table \ref{table:comparison_perceptual_similarity} shows the results. The embedding of \texttt{Message1} results in images with a mean SSIM loss of 8.98\%, while the mean LPIPS loss is 0.29\%. As for \texttt{Message2}, the mean SSIM loss is 6.02\% and the mean LPIPS loss is 0.33\%. The SSIM loss indicates that some human eyes could observe minor effects on the images, but this effect remains small. Moreover, the LPIPS metric reveals that the perturbation due to adversarial embedding is 3 times less noticeable than the ones incurred by JPEG compression. Overall, our results tend to show that the produced adversarial images remain within an acceptable threshold of human perception. 

It is to be noted that the degree of perturbation depends on the choice of the adversarial attack algorithm and its parameter. We selected PGD as a relevant baseline. Still, we can further reduce this impact by lowering the maximum perturbation of PGD or by using alternative algorithms that are known to apply smaller perturbations, e.g. CW \cite{Carlini2017}. In the end, this choice boils down to a compromise between perturbation, efficiency and the rate of success (in creating the adversarial images).

Similarly, using a different classifier leads to different perceptual loss values. In Fig \ref{fig:sampling_similarity}, some models cause lower LPIPS loss when embedding the image while others achieve better performance in terms of SSIM loss. Some models even achieve SSIM loss lower than the threshold a human eye can notice.

\subsection{RQ2: Detection by Steganalysis}
\label{sec:results-detection}


A basic approach in image steganalysis would be to compare the intercepted data with some original reference. This, however, does only work if the difference between the two is only caused by the embedding and not any alteration/noise/tampering during transit.
Steganalysis focuses therefore on identifying the noise profile of different steganography techniques. A basic technique is the LSB embedding: It uses the Least Significant Bits to embed the message to hide.
Other techniques rely either on a statistical analysis of the noise to identify patterns (SPA) or even Deep Neural Network to learn the patterns. The latter technique is the most efficient, it requires, however, the steganalyst to have a large number of labelled images to train the model to recognize the embedding technique. 

We test 2 different detectors on a set of 154 stego-images (generated by encoding Message2) and 154 clean images and measured the AUC score of the detectors:

\begin{itemize}
  \item LSB matching detector: We measure an AUC-ROC score of 0.5, which was expected as this detectors was tailored for steganography techniques that are based on LSB embedding.
  \item SPA dectector: We also measure an AUC-ROC score of 0.5. This however demonstrates that our embedded images do not show common statistical features that can be used to identify them. 
\end{itemize}

To sum up, common detectors do not perform better than random chance into distinguishing relevant pictures (stego-images with Adversarial Embedding) from clean pictures.

\subsection{RQ3: Decoding by Alternative Models}
\label{sec:results-decoding-third-party}

We study whether a different model can classify the adversarial examples (forming the embedding) exactly as the model used to craft them does. We consider \texttt{Message2} and the 154 adversarial images used to encode it, which were produced by applying PGD on the KerasNet. Then we make different models classify the adversarial images: the KerasNet with different parameters, the ResNet model, and the 100 models generated by FeatureNet. All models were trained using the whole Cifar-10 training set. The capability of those models to decode the message is measured as the percentage of images that they classify like the original KerasNet, which we name \emph{decoding rate}.

\begin{figure}
\centering
{\includegraphics[width=\linewidth,trim={4cm 1cm 4cm 3cm},clip]{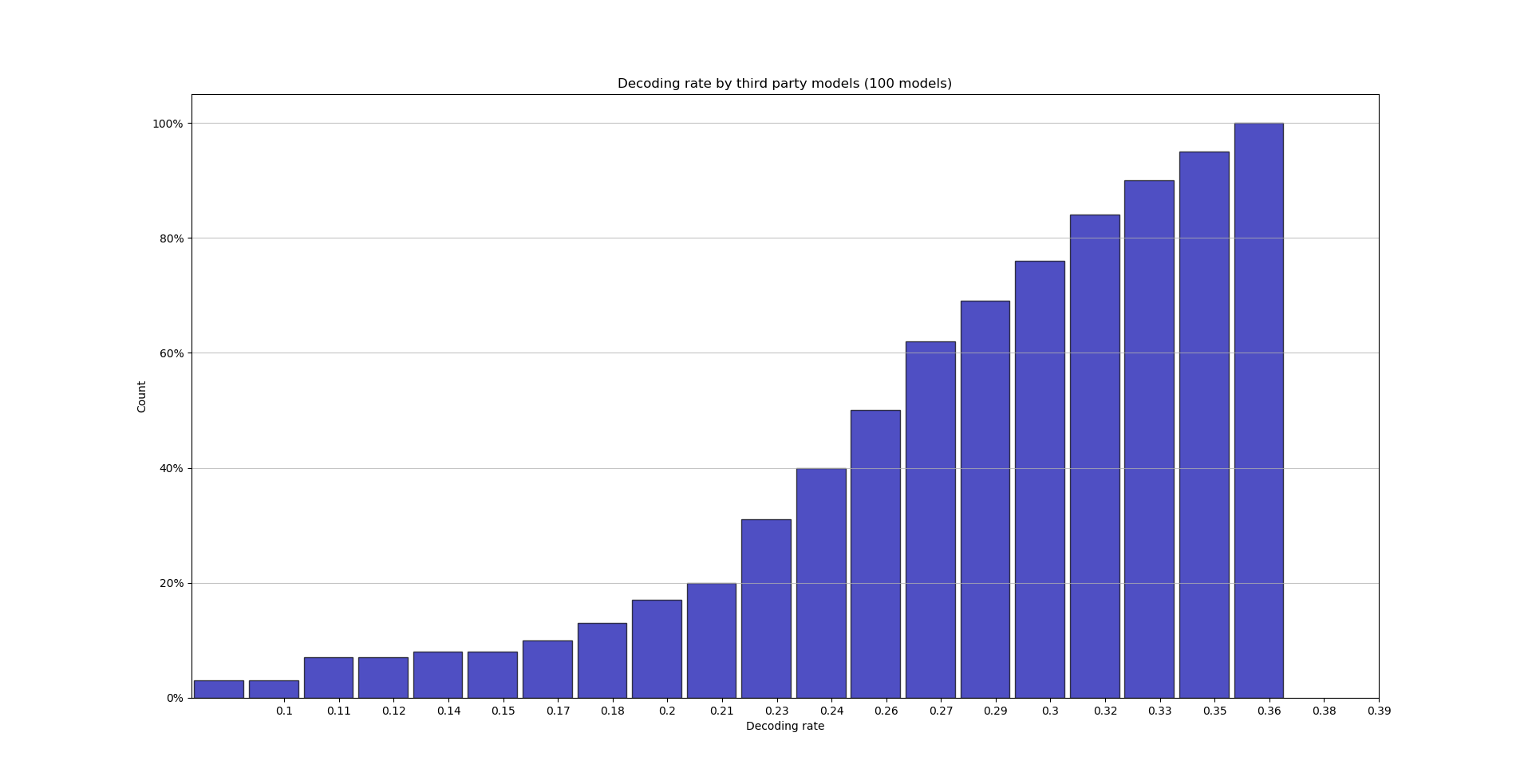}}
\caption{Decoding rate achieved by 100 diverse models generated by FeatureNet.}
\label{fig:decoding-third-party}
\end{figure}

KerasNet with different parameters achieves a decoding rate of 23.72\%, while the ResNet achieves 12.18\%. This indicates that adversarial embedding is highly sensitive to both the model parameters and architecture. To confirm this, we show in Figure \ref{fig:decoding-third-party} the decoding rate achieved by the generated models. It results that no model can retrieve the class (as labelled by the original KerasNet) of more than 37\% of the adversarial images, with more than half failing to surpass 26\% of decoding rate. 

All these low decoding rates increase our confidence that neither randomly-picked models nor handcrafted, state-of-the-art models can break the confidentiality of adversarial-embedded messages. Even if the malicious third party knows the model architecture used for the embedding, differences in parameters also result in a low capability to decode the message illicitly.

\subsection{RQ4: Resilience to image tampering}

\begin{figure}
\centering
\subfloat[Rotation]{\includegraphics[width=\linewidth,trim={4cm 1cm 4cm 1cm},clip]{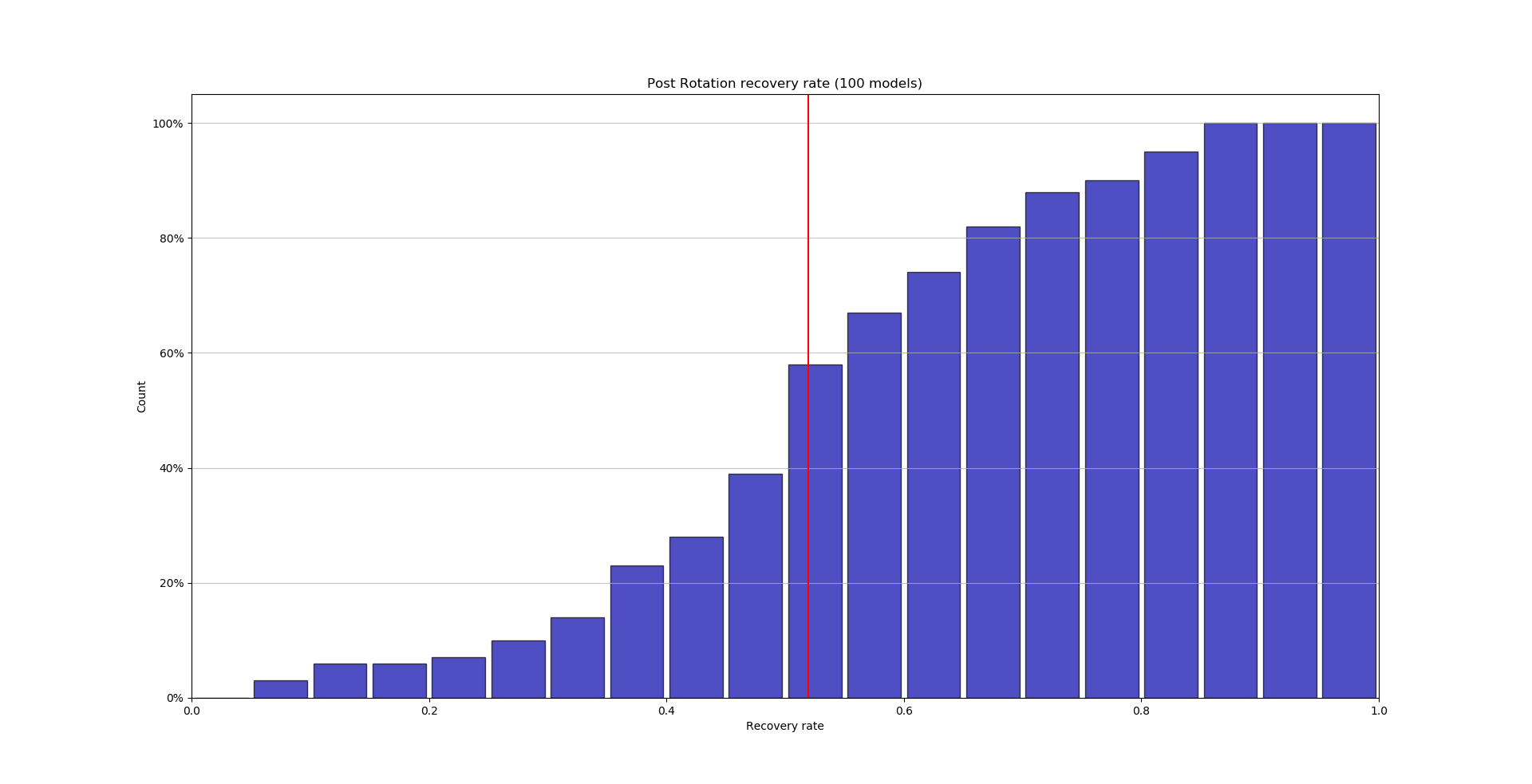}}\\
\subfloat[Upscaling]{\includegraphics[width=\linewidth,trim={4cm 1cm 4cm 1cm},clip]{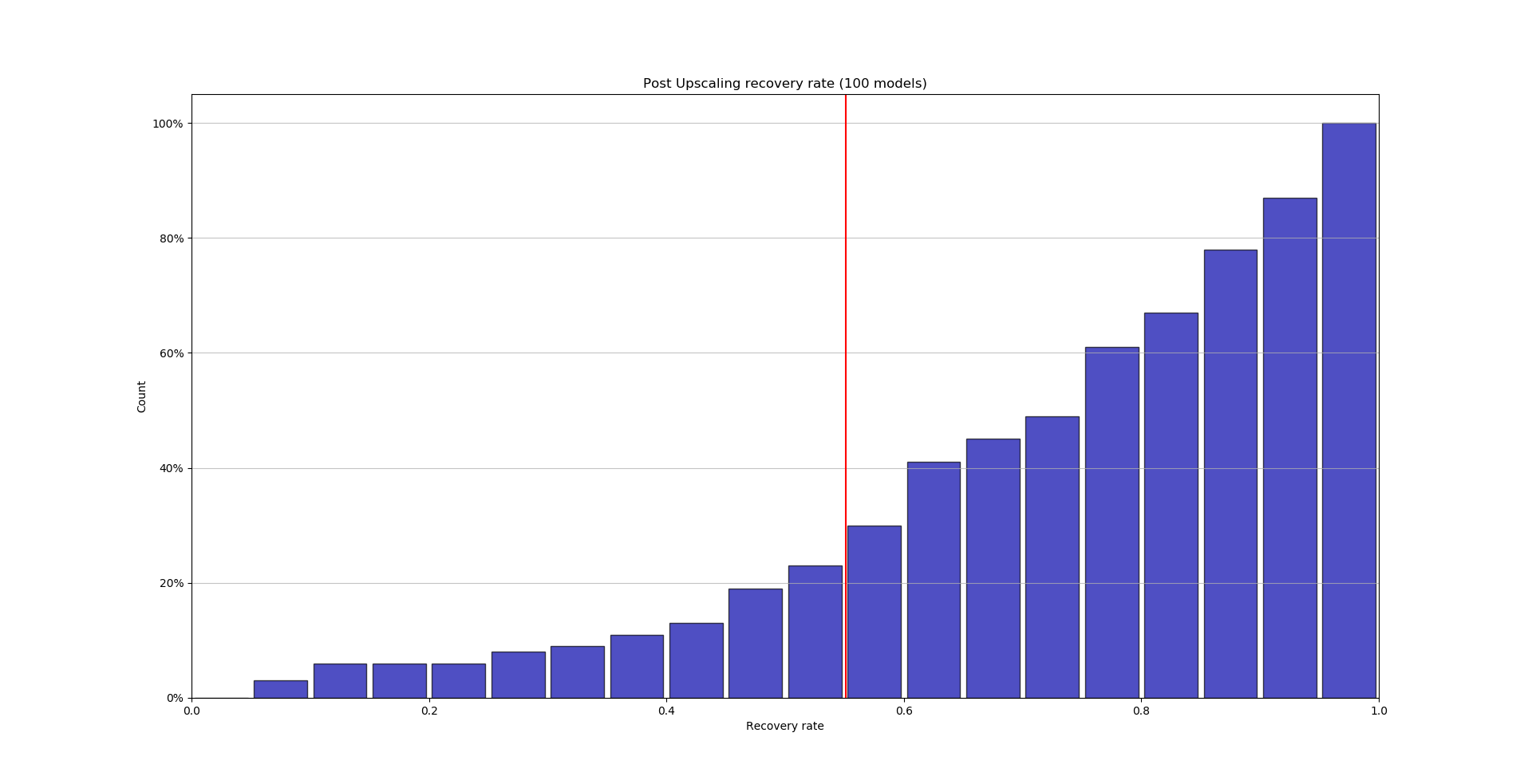}}\\
\subfloat[Cropping]{\includegraphics[width=\linewidth,trim={4cm 1cm 4cm 1cm},clip]{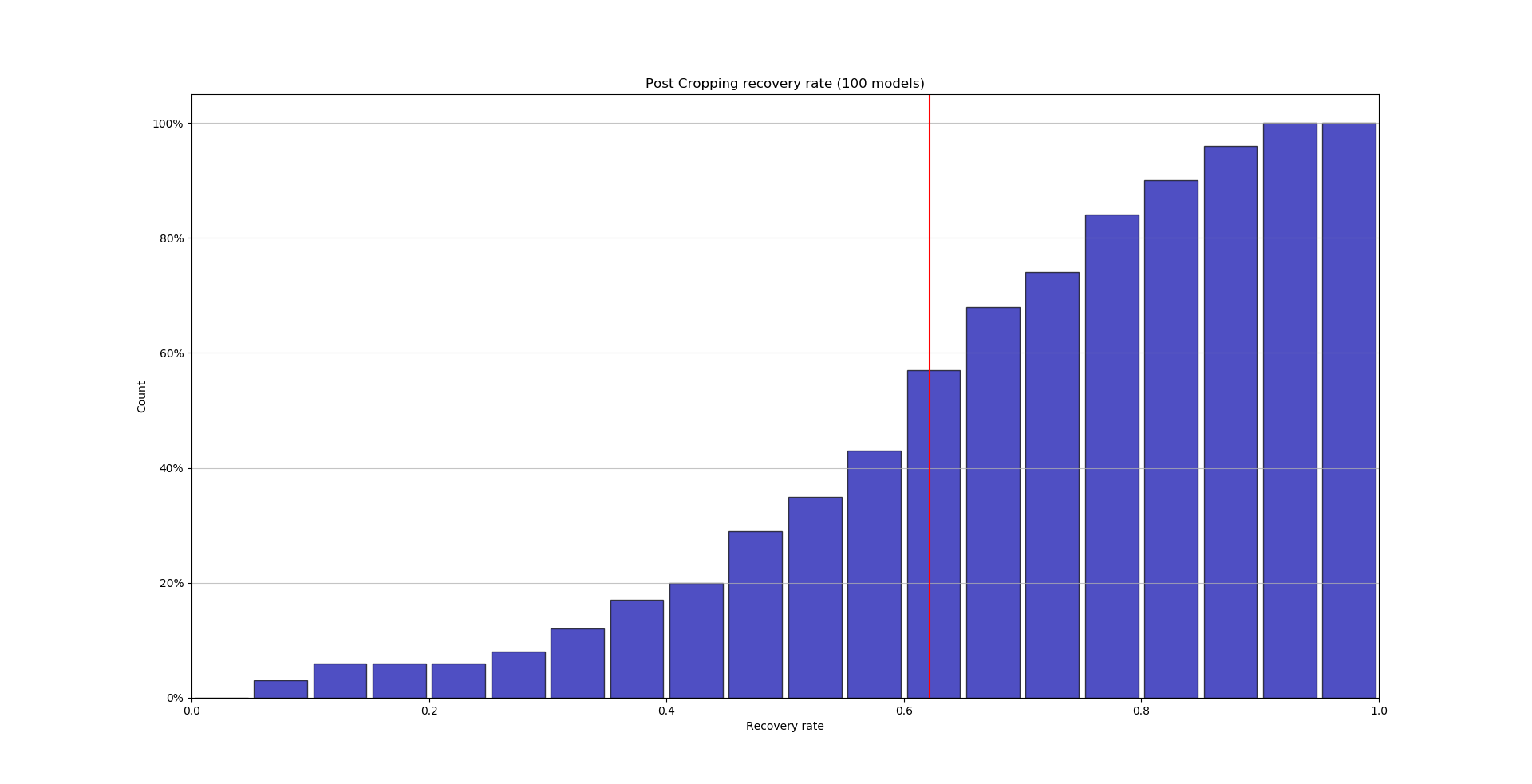}}
\caption{Recovery rate of the 100 generated models against three spatial-transformations of the stego-images. The red lines indicate the values we obtained by KerasNet model.
Regardless of the transformation, there is always a model that can be used to achieve a high embedding recovery. We achieve up to 94.9\% recovery rate after rotation perturbation, 94.2\% after cropping and 99.4\% recovery rate after upscaling. This confirms the high flexibility of our approach and demonstrates its high resilience against spatial-based transformation.}
\label{fig:spatial-tampering}
\end{figure}

\subsubsection{Spatial Domain Tampering}

We focus first on soft image tampering and three local image alterations: 
\begin{itemize}
  \item Rotation: We rotate the images by 15°.  
  \item Upscaling: We use bilinear interpolation to resize the images to 64x64 pixels.  
  \item Cropping: We remove 12.5\% of the images by cropping, keeping only the central part.
\end{itemize}
These transformations are common when copyrighted images are shared illegally \cite{podilchuk2001digital}.

To measure the resilience of adversarial embedding to those transformations, we consider the 100 generated models and the images. We create three altered versions of each image, using the three transformations above independently. Then, for each model $m$, original image $o$ and altered image $a$, we check whether $m$ assigns the same class to $o$ and $a$. If that is not the case, it means that the alteration results in changing the classification of the image, thereby threatening the integrity of messages encoded by adversarial embedding. We measure the resilience of $m$ to each transformation $t$ by computing the recovery rate of $m$ against $t$, i.e. the percentage of images resulting by applying $t$ that $m$ classifies in the same class as their original counterpart.

Figure \ref{fig:spatial-tampering} shows, for each transformation, the recovery rates achieved by the 100 models. We observe that, for all transformations, we can always find a classification model that achieves a high recovery rate. 99\% for upscaling transformation, 94.9\% for rotation transformation and 94.2\% under cropping perturbation, these results indicate that adversarial embedding is resilient to spatial-tampering of the images and that we can craft adversarial models strong against such transformations.

It is worth noticing that our default model, KerasNet, crafted and trained to achieve high classification accuracy is not the best model to ensure the most robust embedding.





\subsubsection{Frequency Domain perturbations}

\begin{figure}
\centering
\subfloat[Low JPEG compression (90\%)]{\includegraphics[width=\linewidth,trim={4cm 1cm 4cm 1cm},clip]{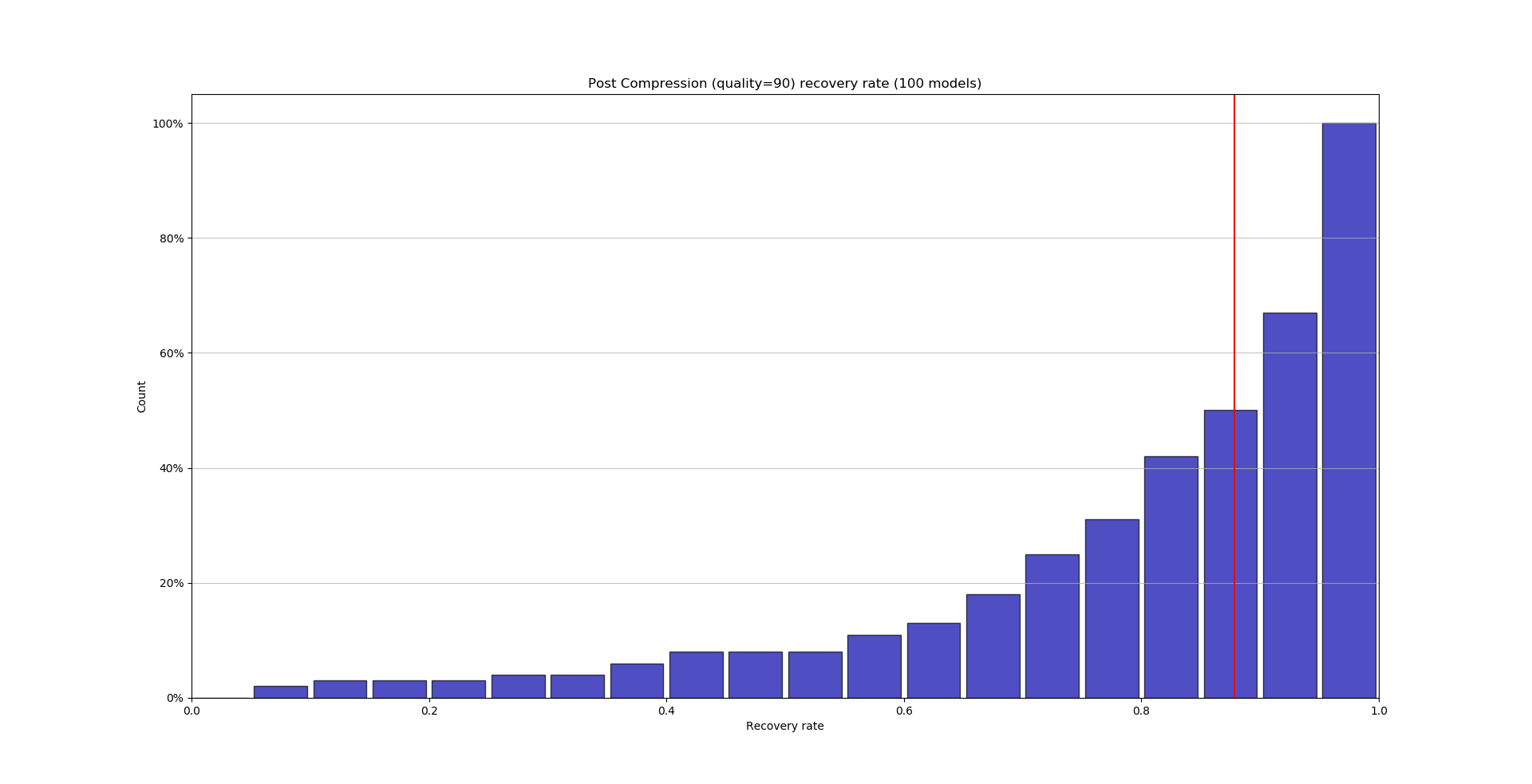}}\\
\subfloat[Average JPEG compression (75\%)]{\includegraphics[width=\linewidth,trim={4cm 1cm 4cm 1cm},clip]{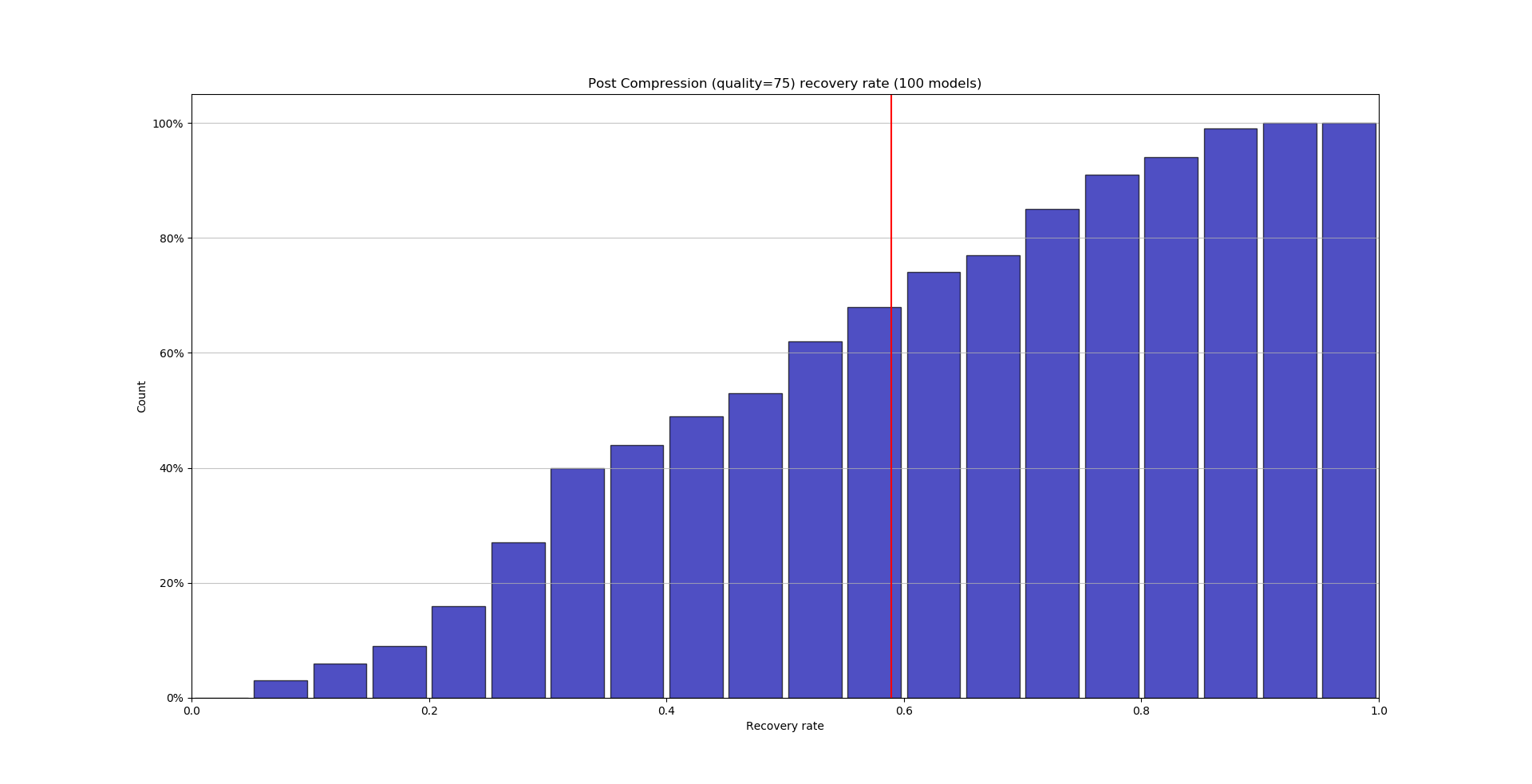}}\\
\subfloat[High JPEG compression (50\%)]{\includegraphics[width=\linewidth,trim={4cm 1cm 4cm 1cm},clip]{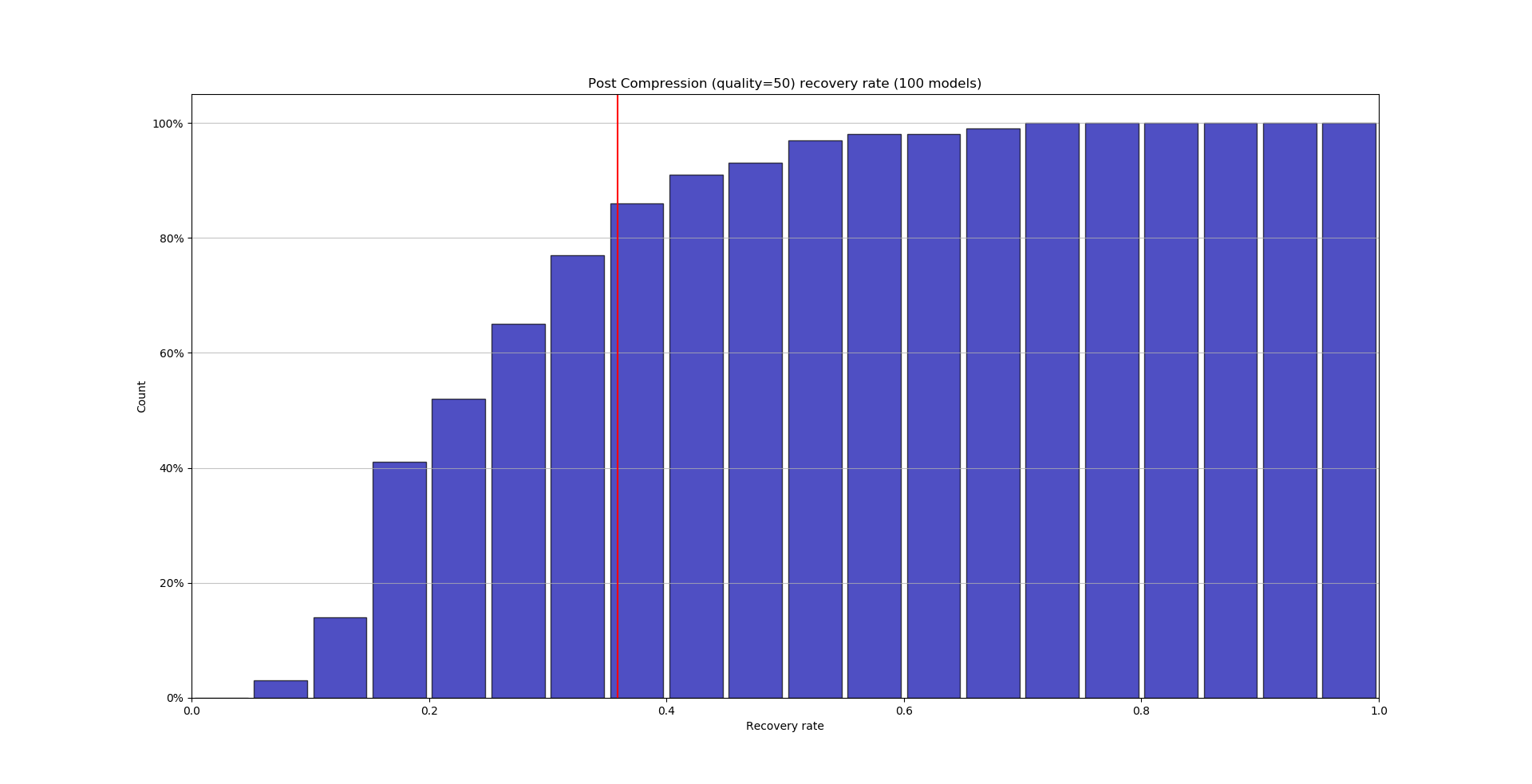}}\\
\subfloat[Color Depth Reduction]{\includegraphics[width=\linewidth,trim={4cm 1cm 4cm 1cm},clip]{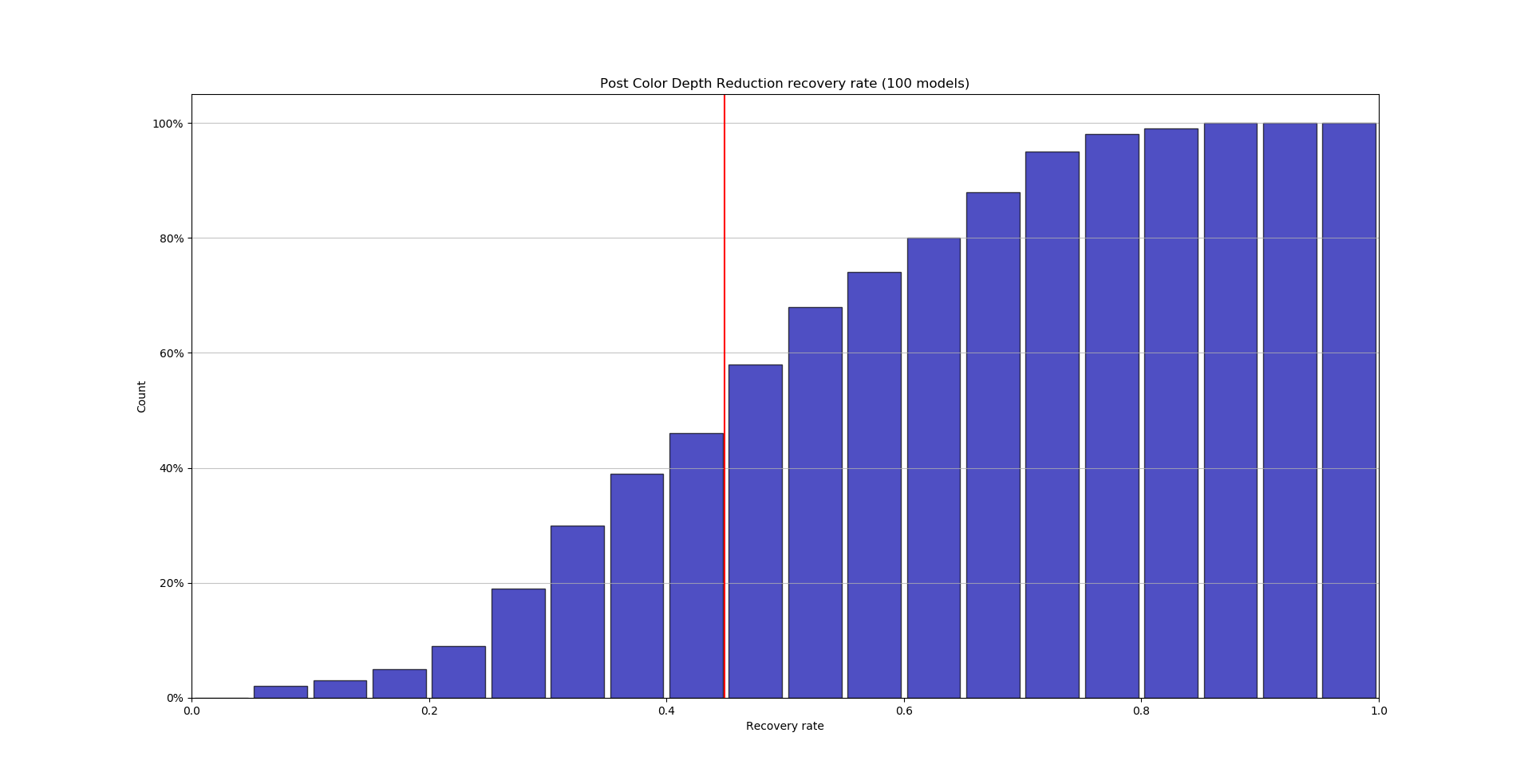}}\\
\caption{Recovery rate of the 100 generated models against aggressive image transformations. The red lines indicate the values we obtained by the KerasNet model we studied in details.}
\label{fig:aggressive-tempering}
\end{figure}

Next, we study the impact of two aggressive image tampering: JPEG compression (A frequency domain transformation) and Color Depth Reduction (CDR).
JPEG compression relies on various steps (color transformation, DCT, quantization) that cause information loss.
CDR reduces the number of bits used to encode different colors. For instance, Cifar-10 images use a 32 bits color depth. This means that every channel is encoded on 32bits. Reducing color depth to 8 bits makes the picture contain less tone variation and fewer details. We apply JPEG compression with a 90\%, a 75\% and 50\% quality rates (resulting in loss of information of 10\%, 25\% and 50\%, respectively) and CDR (to 8 bits, leading to pictures with only 1/12 of the original information) independently and we measure again the recovery rate achieved by the 100 models.

In Fig \ref{fig:aggressive-tempering}(a), our models achieve up to 100\% recovery rate under jpeg compression (Q=90). When the compression rate increases, we can still achieve up to 95\% (Fig \ref{fig:aggressive-tempering}(b)) recovery rate against jpeg(Q=75) and one of our models achieves more than 72\% recovery rate under a jpeg(Q=50) compression(\ref{fig:aggressive-tempering}(c)). 

Our models also show a spread robustness to color depth reduction (Fig \ref{fig:aggressive-tempering}(d)) and reach up to 88\% recovery rate.

\subsection{RQ5: SATA Embedding}
\label{sec:sata_evaluation}





\subsubsection{\textbf{Embedding Capacity}}
\label{subsection:embedding_capacity_experiments}

To determine how much data our adversarial embedding with SATA could achieve, we randomly pick 1,000 messages of 6,643 bits and use our technique to determine how many pictures are needed depending on how much classes we would embed per picture. 
Once we know the number of pictures needed, we measure the embedding density in the case of color pictures of 32x32 pixels. 

\begin{figure}
\centering
\subfloat[Embedding Density for classifiers with 10 classes]{\includegraphics[width=\linewidth,trim={4cm 1cm 4cm 0.85cm},clip]{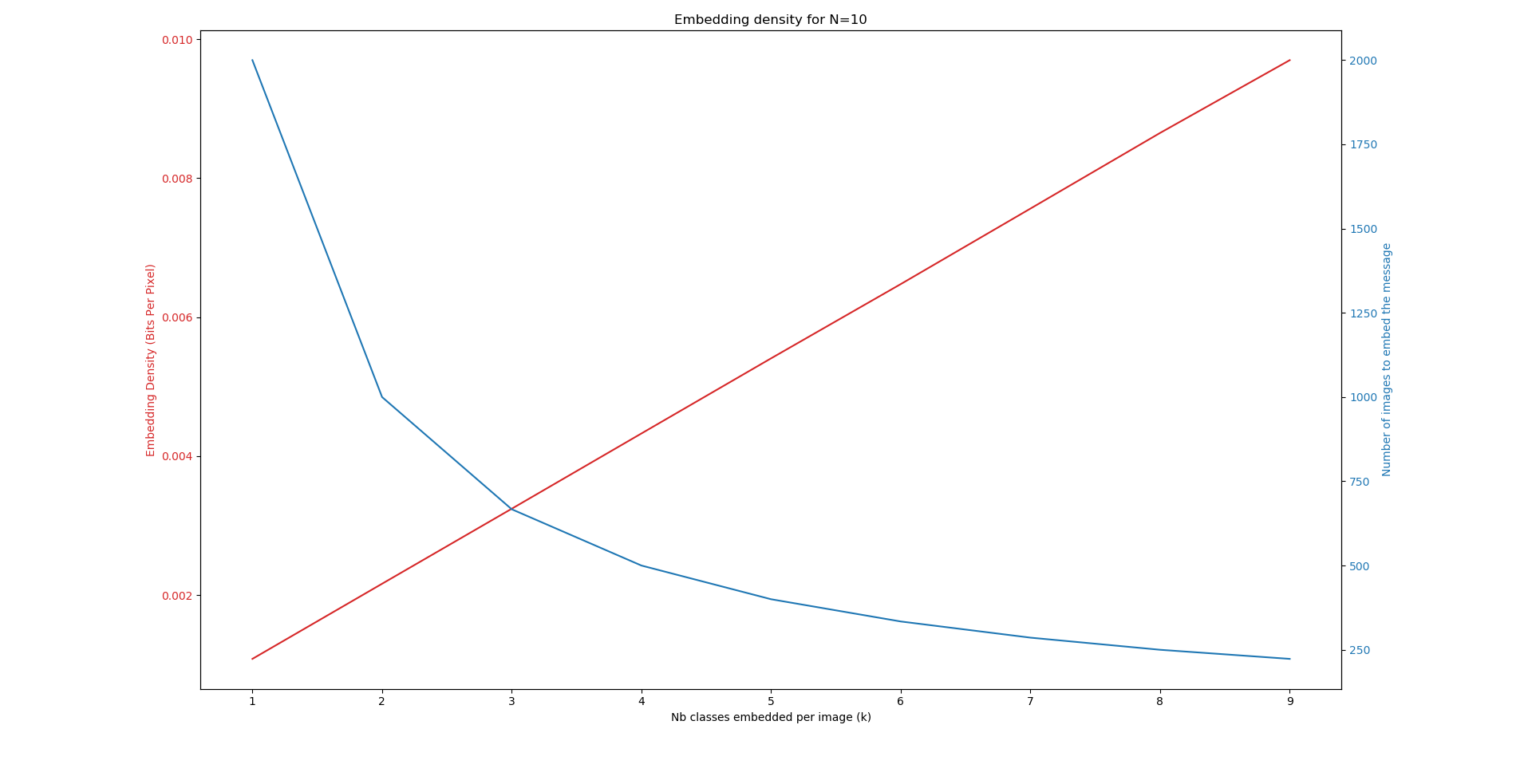}}\\
\subfloat[Embedding Density for classifiers with 100 classes]{\includegraphics[width=\linewidth,trim={4cm 1cm 4cm 0.85cm},clip]{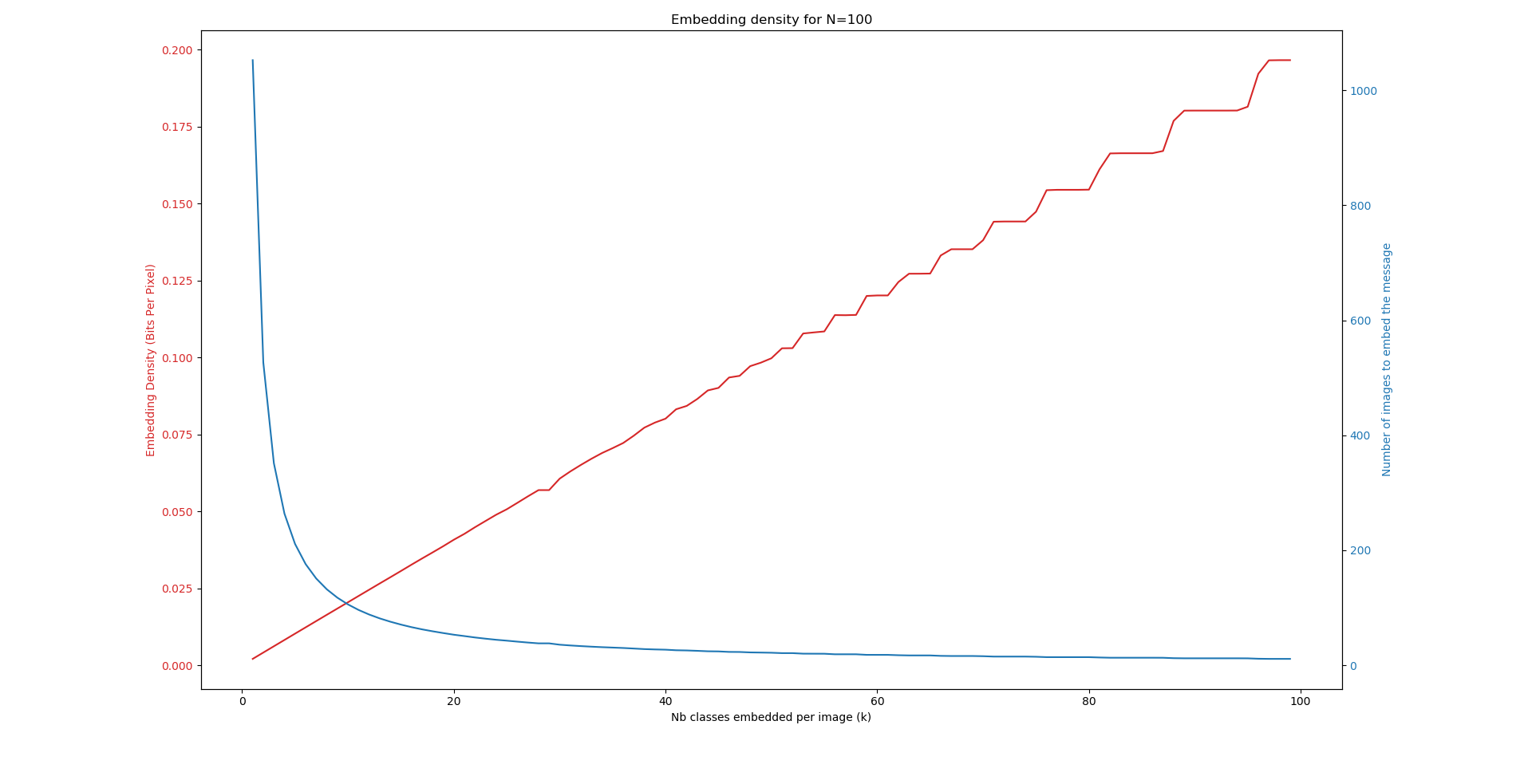}}
\caption{Average number of images used to embed 1,000 random messages of 6,643 bits each (blue line) and its associated embedding density (red line) as we increase the number of classes embedded per picturs using  10 and 100 classes classification models.}
\label{fig:embedding_capacity_sata}
\end{figure}

Fig \ref{fig:embedding_capacity_sata}(a) confirms our previous estimate that a 10-class classifiers has a density from $1.87e^-3$ BPP with 2 classes embedded per image to $9.7e^-3$ BPP when embedding 9 classes per image.

With a 100-classes image classification model (Fig \ref{fig:embedding_capacity_sata}(b)), we can achieve up to $0.2$ BPP (standard density of steganography techniques \cite{zhu2018hidden}).

\begin{figure}
\centering
\includegraphics[width=\linewidth,trim={4cm 1cm 4cm 0.85cm},clip]{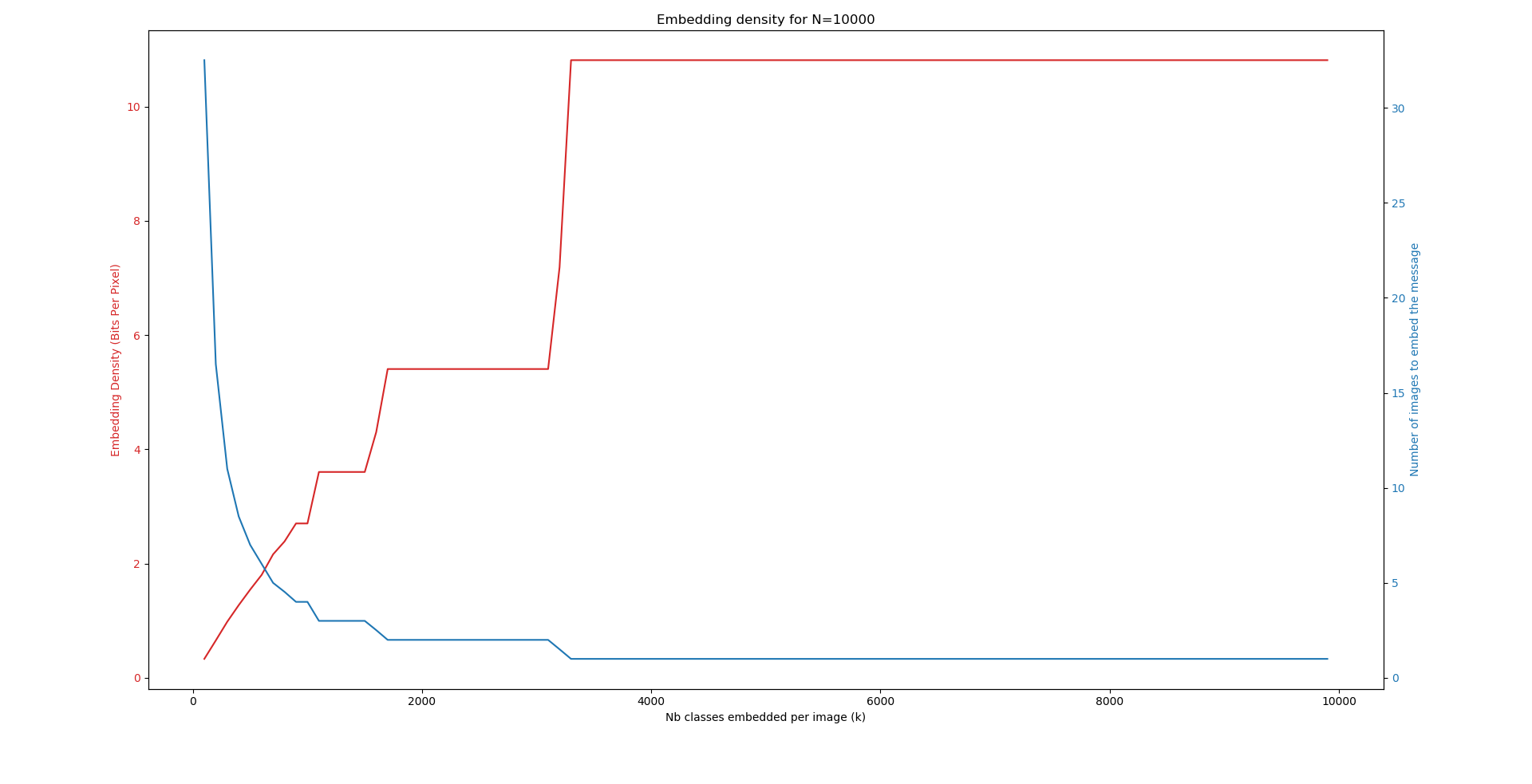}
\caption{We embed 1,000 messages of 33 Kbits eachs using a 10,000 classes model. The plot tracks the average number of images used (blue line) and its associated embedding density (red line) as we increase the number of classes embedded per picture.}
\label{fig:embedding_capacity_sata_extreme}
\end{figure}

If we use larger models that support 10.000 classes (ImageNet classifiers for instance), we exceed a density of 10 Bits per Pixel (Fig \ref{fig:embedding_capacity_sata_extreme}), improving over any existing steganography technique \cite{stegano_survey}. 

\subsubsection{\textbf{Embedding Capacity and Success Rate Trade-off}}

To embed many classes per image, we have to tweak the hyperparameters of our embedding, to ensure that we recover all the classes encoded in the right order (i.e. the success rate of our embedding).

\label{subsection:embedding_capacity_tradeoff}
To maximize the success rate, we used a grid search to find the best combination of the following two SATA hyperparameters:
\begin{itemize}
  \item $\gamma$, which controls the relative weights of the $k$ forced classes.
  \item $\epsilon$ the maximum amount of perturbation (measured as L2 distance) SATA can apply to an image.
\end{itemize}
The best $\epsilon$ was always 0.5, while the ideal value of $\gamma$ changed depending on $k$. 

SATA attack has also other hyperparameters that we kept unchanged from the PGD implementation. For instance the perturbation step $ \epsilon_\text{step}=0.1$  and the number of random initialisations within the epsilon ball $num_\text{random\_init}=10$.

We ran our experiment using the KerasNet model and picked our cover images randomly from the cifar-10 dataset. 

\begin{figure}
\centering
\includegraphics[width=8cm]{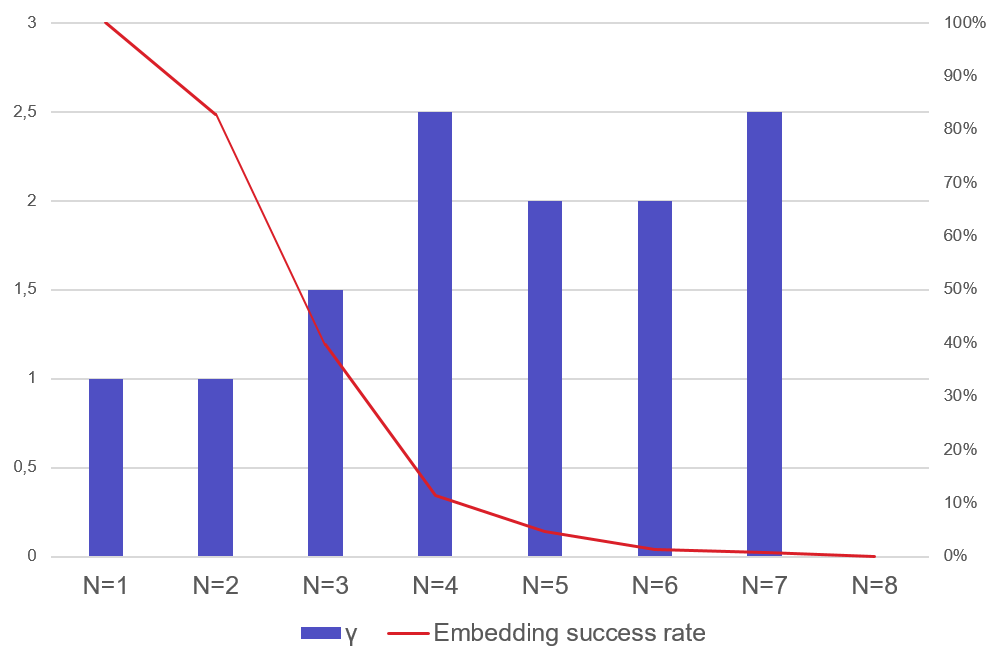}
\caption{Best SATA Multi-class embedding hyper-parameter $\gamma$ values and their associate mean embedding sucess rate over 1000 cover images.}
\label{fig:sata_embedding}
\end{figure}

We associate each image with a binary result set to \emph{success} SATA managed to find a successful perturbation leading to the intended encoding; otherwise, it is set to \emph{failed}. Then, the success rate of SATA is measured as the percentage of images for which it was successful.

Table \ref{fig:sata_embedding} presents, for values of $k$, the success rate when SATA embeds a randomly-generated message of 6,643 bits. 
We see that SATA can always find at least an image to embed up to seven classes per-image. Eight classes or more are not achievable using our current model.  For $k = 7$, the success rate over 1000 images was 0.8\%, meaning that SATA was successful for 8 images. 






\section{Conclusion}
\label{sec:conclusion}

In this paper, we proposed an embedding pipeline that hides secret messages in images by using image classification (DNN) models and adversarial attack algorithms. We have shown that such an approach can be used to form a highly elusive (bypass detection techniques), flexible and customizable steganography and watermarking technique. We have also demonstrated that the combined use of a 1,000 output class model with our new sorted adversarial attack algorithm can achieve high-density embeddings (higher than existing steganography research). We also showed that our embeddings are resilient to image tampering, e.g., jpeg compression.  

An inherent benefit of our approach is that it leverages adversarial attack algorithms in a black-box way. Therefore, our technique can take advantage of any, current or future, state-of-the-art technique (targeted adversarial) coming out from this highly active research area and (theoretically) can be at least as effective as existing adversarial attacks (our technique will remain effective as long as adversarial attacks manage to remain undetected).   




Future work should attempt to expand our study to larger models and datasets, 
including other media where adversarial examples have showed mature results (such as audio, video and text).  


\bibliographystyle{IEEEtran}
\bibliography{bib/various,bib/attacks,bib/defense,bib/adv_applications,bib/stegano,bib/detection}
\end{document}